\documentclass[journal,twoside]{IEEEtran}
\usepackage{multirow}
\usepackage{makecell}
\usepackage{amsmath}
\usepackage{amsthm}
\usepackage{amssymb}
\usepackage{graphicx}
\usepackage[noend]{algpseudocode}
\usepackage{algorithm, algorithmicx}
\usepackage{threeparttable}
\usepackage{color}
\usepackage{booktabs}
\UseRawInputEncoding
\usepackage[bookmarks=false]{hyperref}
\newtheorem{definition}{Definition}

\hypersetup{hidelinks}
\ifCLASSINFOpdf
\else
\fi
\begin{document}
	
\title{Model Checking of Workflow Nets with Tables and Constraints}

\author{Jian Song\, and Guanjun Liu\,~\IEEEmembership{Senior Member,~IEEE}
\thanks{~J.~Song and G.~J.~Liu are with the Department of Computer Science and Technology, Tongji University, Shanghai 201804, China (e-mail: 1910690@tongji.edu.cn; liuguanjun@tongji.edu.cn).}
}


\maketitle

\begin{abstract}
 Many operations in workflow systems are dependent on database tables. The classical workflow net and its extensions (e.g., workflow net with data) cannot model these operations, so that they cannot find some related errors. Recently, workflow nets with tables (WFT-nets) were proposed to remedy such a flaw. However, when the reachability graph of a WFT-net is constructed by their method, some pseudo states can be generated since it does not consider the guards that constrain the enabling and firing of transitions. Additionally, only the soundness property of WFT-nets is considered that represents a single design requirement, while many other requirements, especially those related to tables, cannot be analyzed. In this paper, we re-define the WFT-net by augmenting the constraints of guards to it and re-name it as workflow net with tables and constraints (WFTC-net). We propose a new method to generate the state reachability graph (SRG) of a WFTC-net such that the SRG can avoid pseudo states, due to the consideration of the guards in it. To represent design requirements related to database operations, such as design requirements and data flow errors, we define database-oriented computation tree logic (DCTL). We design the model checking algorithms of DCTL based on the SRG of WFTC-nets and develop a tool. Experiments on several public benchmarks show the usefulness of our methods.
		
	\end{abstract}
	
	
	\IEEEpeerreviewmaketitle
	
	\section{Introduction}
	\IEEEPARstart{W}{ith} the development of information technology and service computing, workflow systems have become increasingly popular and played an essential role in e-commerce and intelligent medical enterprises \cite{ref67,ref68}. Some errors often occur in workflow systems since they are increasingly complex. If these errors cannot be found before implementing a workflow system, some serious results are possibly caused. The formal modeling and analysis of workflow systems have become an important way to find these errors \cite{ref9,ref64,ref70}.
	
	As a formal language, Petri nets (PN) are widely used in modeling and analyzing workflow systems \cite{ref10,ref11,ref69}. Using the structural characteristics of PN can effectively find some control flow errors \cite{ref7}. Although the traditional workflow net (WF-net) can simulate the control flow of activities \cite{ref1,ref2,ref3} and analyze control flow errors \cite{ref12,ref13}, it cannot describe the data operations so that some errors related to them cannot be checked by this class of PN. A data flow reflects the dependence of data items in a workflow system \cite{ref4,ref5}. Unreasonable data operations can still lead to many errors, such as data inconsistency and deadlock \cite{ref7,ref8}.
	
	In order to make up for the lack of data description in WF-nets, Sidorova et al. propose workflow net with data (WFD-net) which extends WF-net with conceptualized data operations (i.e., $read$, $write$ and $delete$) \cite{ref14,ref15,ref16}. Combi et al. propose another conceptual modeling method of workflow and data integration in order to detect the possible defects of data, using the concept of activity view to capture the data operations and providing a more rigorous analysis method \cite{ref18,ref5,ref20}. Meda et al. use WFD-net to detect data flow errors through anti-patterns \cite{ref33,ref34}. When workflow systems become larger and more complex, using data refinement methods can check data flow errors and alleviate the state explosion problem \cite{ref21,ref22}.
	
	In a workflow system, many operations depend on database tables. Some data-related errors and logical defects are related to these operations, but they cannot be reflected by the above models. Therefore, we propose workflow net with tables (WFT-net) \cite{ref39} in which the operation statements of database tables are bound to transitions, establishing a relationship between a control flow and database operations. WFT-nets can be used to detect some control flow flaws via representing the changes of tables, while these flaws cannot be detected by WF-nets or WFD-nets \cite{ref39}.
	
	A coarse-grained data refinement method is proposed to generate the state reachability graph (SRG) of a WFT-net in \cite{ref39}. However, because the constraint and dependence among guards in a WFT-net are not considered in this refinement method, pseudo states are possibly produced in an SRG. In this paper, we re-define WFT-net named workflow net with tables and constraints (WFTC-net) in order to describe these constraints more accurately. Our method of generating the SRG of a WFTC-net can effectively avoid pseudo states. Additionally, only one property, namely soundness, is considered in \cite{ref39}. Soundness as a basic design requirement guarantees that a final state is always reachable. However, many other design requirements related to tables cannot be represented by soundness. Therefore, in order to represent more design requirements over WFTC-nets, we propose database-oriented computation tree logic (DCTL) in this paper.
	
Model checking is an automatical technique used to verify the correctness of workflow systems \cite{ref35,ref29,ref30}. A temporal logic language such as linear temporal logic (LTL) and computation tree logic (CTL) is usually used to specify the design requirements of a system \cite{ref27,ref37,ref61,ref66}. A computer searches the state space of a model to confirm if a logic formula is true or false over the model. Generally, if the formula is true, then the system satisfies the related requirement or does not satisfy the requirement (a counterexample is output) \cite{ref25,ref26,ref36,ref52}. Our DCTL considers the descriptions of database tables as atomic propositions so that they can represent some data-related design requirements. We propose model checking algorithms and develop a tool. We also do a number of experiments on a group of benchmarks that illustrate the usefulness of our model and methods.

The rest of this paper is organized as follows. Section~\uppercase\expandafter{\romannumeral2} presents some basic notions and an example of motivation. Section~\uppercase\expandafter{\romannumeral3} defines WFTC-net and its firing rules and proposes a data refinement method used to generate the SRG of a WFTC-net. Section~\uppercase\expandafter{\romannumeral4} presents DCTL and the related model checking algorithm. Section~\uppercase\expandafter{\romannumeral5} evaluates our tool and a group of experiments. Section~\uppercase\expandafter{\romannumeral6} concludes this paper.	
	
\section{Basic Notions and Motivating Example}
\subsection{Basic Notions}
A net is a 3-tuple $N=(P$, $T$, $F)$, where $P$ is a finite set of places, $T$ is a finite set of transitions, $F\subseteq(P\times T)\cup (T\times P)$ is a flow relation, and $P \cap T = \emptyset \wedge P \cup T \ne \emptyset$. A marking of a net is a mapping $m: P\rightarrow \mathbb{N}$, where $\mathbb{N}$ is the set of non-negative integers and $m(p)$ is the number of tokens in place $p$.  For each node $x\in P\cup T$, its preset is denoted by ${}^ \bullet x = \left\{ {y\left| {y \in P \cup T \wedge \left( {y,x} \right) \in F} \right.} \right\}$, and its postset is denoted by ${x^ \bullet } = \left\{ {y\left| {y \in P \cup T \wedge \left( {x,y} \right) \in F} \right.} \right\}$.	
	
\begin{definition}[Petri net and firing rule \cite{ref40,ref41}]
	A net $N$ with an \emph{initial marking} $m_0$ is a \emph{Petri net} and denoted as $PN=(N$, $m_0)$. If $\forall p \in P:p \in {}^ \bullet t \to m\left( p \right) \geq 1$, then transition $t$ is enabled at marking $m$, which is denoted by $m[t\rangle$. Firing an enabled transition $t$ at marking $m$ yields a new marking $m^\prime$, which is denoted as $m[t\rangle m'$, where $\forall p \in P$:
	
	$
	\begin{centering}
		m^\prime \left( p \right) = \left\{ \begin{array}{l}
			m\left( p \right) - 1,{\rm{  }} \quad if \quad {\rm{ }}p \in {}^ \bullet t - {t^ \bullet };\\
			m\left( p \right) + 1,{\rm{ }} \quad if \quad {\rm{ }}p \in {t^ \bullet } - {}^ \bullet t;\\
			m\left( p \right),{\rm{   }}  \quad\quad\,\,\, otherwise.
		\end{array} \right.
	\end{centering}
	$
	\label{defn1}
\end{definition}	
	
A marking $m'$ is reachable from $m$, if there is a firing transition sequence $t_1$, $t_2$, ..., $t_k$ and the marking sequence $m_1$, $m_2$, ..., $m_k$ such that $m[t_1\rangle m_1[t_2\rangle m_2...m_{k-1}[t_k\rangle m_k$. The set of all marking reachable from $m$ is denoted as $R(m)$. If the transition sequence $\sigma=t_1, t_2, ..., t_k$, then the above formula can be denoted as $m[\sigma\rangle m'$. 

\begin{definition}[Workflow nets \cite{ref42,ref43}]
	A net $N=(P$, $T$, $F)$ is a workflow net (WF-net) if:
	\begin{itemize}
		\item[1.] $N$ has two special places, $i.e.,$ one source place \emph{start} and one sink place \emph{end} in $P$ such that $^\bullet$\emph{start} $= \emptyset$ and \emph{end}$^\bullet = \emptyset;$ and
		\item[2.] $\forall$ $x\in P\cup T:$ $(\emph{start},$ $x)\in F^*$ and $(x,$ \emph{end}$)\in F^*,$ where $F^*$ is the reflexive-transitive closure of $F.$
	\end{itemize}
	\label{defn2}
\end{definition}

In order to model the control flow and data flow in a workflow system, workflow net with data (WFD-net) is formed by adding data operations and guards to a WF-net.
	
\begin{definition}[Workflow nets with data \cite{ref15,ref23}]
	A WFD-net is an 8-tuple  $N = (P,$ $T,$ $F,$ $D,$ $rd,$ $wt,$ $dt,$ $grd)$ where
	\begin{itemize}
		\item[(1)] $(P,$ $T,$ $F)$ is a WF-net;
		\item[(2)] $D$ is a finite set of data items;
		\item[(3)] $rd:T \to {2^D}$ is a labeling function of reading data;
		\item[(4)] $wt:T \to {2^D}$ is a labeling function of writing data;
		\item[(5)] $dt:T \to {2^D}$ is a labeling function of deleting data; and
		\item[(6)] $grd:T \to G_\Pi$ is a labeling function of assigning a guard for each transition. $G_\Pi$ is a set of guards, while each guard is a Boolean expression over a set of predicates $\Pi=\{\pi_1$, $\pi_2$, $...$, $\pi_n\}$. $\varGamma:\Pi \to {2^D}$ is a predicate labeling function, and it assigns the dependent data items for each predicate. When $\varGamma$$(\pi)=\{d_1,...,d_n\}$, we use $\pi(d_1,...,d_n)$ to represent. When performing a data operation on a data item $d_i$, the predicate $\pi \to \{T,F,\bot\}$ has three possible values: true, false, and undefined.
	\end{itemize}	
	\label{defn3}
\end{definition}	
	
Figure \ref{fig02} (a) shows a WFD-net of a Vehicle Management. The explanation of the predicate expressions and guards are shown in Figure \ref{fig02} (b). Its data item set is $D$ = $\{id$, $password$, $license$, $copy\}$ and its predicate set is $\Pi$ = $\{\pi_1=registered(id)$, $\pi_2=exist(license)$, $\pi_3=true(copy)\}$. For example, predicate $\pi_1$ = $registered(id)$ depends on the value of $id$: if the value of the data item $id$ belongs to a registered $user$, then the value of the predicate $\pi_1$ is $true$, and otherwise it is $false$. However, because the data item $id$ is independent of the predicates $\pi_2$ and $\pi_3$, we have that $\pi_2=\bot$ and $\pi_3=\bot$. Guard set is $G_\Pi$ = $\{ g_1$, $g_2$, $g_3$, $g_4$, $g_5$, $g_6\}$, $grd(t_1)$ = $g_1$ = $\pi_1$, $grd(t_2)$ = $g_2=\neg\pi_1$, etc. In function $wt$, $wt(t_0)$ = $\{id$, $password\}$ means that a $write$ operation on data items $id$ and password are performed when firing $t_0$. In function $rd$, $rd(t_{13})$ = $\{license\}$ means that a $read$ operation on a data item $license$ is performed when firing $t_{13}$. Other labels can be understood similarly.	
	
\begin{definition}[Table]
	A table $R=\{r_1, r_2,..., r_n\}$ is a finite set of records. A record $r_i=\{d_1, d_2,..., d_k\}$ represents the values of  $k$ attributes, where $d_i$ represents the value of the $i$-th attribute value for each i $\in$ $\{1,...,k\}$.
	\label{defn5}
\end{definition}

For example, a table $User$ is shown in Figure~\ref{fig06} (b), which contains two records $r_1=\{id1, license1, copy1\}$ and $r_2=\{id2, license2, copy2\}$.

\begin{definition}[Workflow Net with Table \cite{ref39}]
	A workflow net with table (WFT-net) is a 14-tuple $N = (P$, $T$, $F$, $G$, $D$, $R$, $rd$, $wt$, $dt$, $sel$, $ins$, $del$, $upd$, $guard)$ where
	\begin{itemize}
		\item[(1)] $(P$, $T$, $F)$ is a WF-net;
		\item[(2)] $G$ is a set of guards, $var(G)$ represents the variables in the guards $G$;
		\item[(3)] $D$ is a finite set of data items;
		\item[(4)] $R=\{r_1$, $r_2$, $...$, $r_n\}$ is an initial table consisting of $n$ records; The data in D can be included in R or not. Generally, D describes an abstract data operation, and mapping to R is a concrete data item value.
		\item[(5)] $rd:T \to {2^D}$ is a labeling function of reading data;
		\item[(6)] $wt:T \to {2^D}$ is a labeling function of writing data;
		\item[(7)] $dt:T \to {2^D}$ is a labeling function of deleting data;
		\item[(8)] $sel:T \to {2^R}$ is a labeling function of selecting a record in a table;
		\item[(9)] $ins:T \to {2^R}$ is a labeling function of inserting a record in a table;
		\item[(10)] $del:T \to {2^R}$ is a labeling function of deleting a record in a table;
		\item[(11)] $upd:T \to {2^R}$ is a labeling function of updating a record in a table; and
		\item[(12)] $guard:T \to G_\Pi$ is a labeling function of guards. ~$G_\Pi$ is a set of guards, each of which is a Boolean expression over a set of predicates $\Pi=\{\pi_1$, $\pi_2$, $...$, $\pi_n\}$ where $\pi_i$ is a predicate defined on $D$ or $R$. $\varGamma:\Pi \to {2^{D\cup R}}$ is a predicate labeling function, and it assigns the dependent data items for each predicate. When $\varGamma$$(\pi)=\{d_1,...,d_n\}$, we use $\pi(d_1,...,d_n)$ to represent. When performing a data operation on a data item $d_i$, the predicate $\pi \to \{T,F,\bot\}$ has three possible values: true, false, and undefined.
	\end{itemize}	
	\label{defn44}
\end{definition}

The firing rules of WFD-net and WFT-net are described in \cite{ref15,ref16,ref39}. This article will not introduce them too much, and they can be understood according to the firing rules of our WFTC-net. Notice that $sel$, $ins$, $del$, and $upd$ are used to simulate the related operations in the relation database.

\subsection{A motivating example}

	\begin{figure} [tt]
	\centering
	\includegraphics[width=0.48\textwidth]{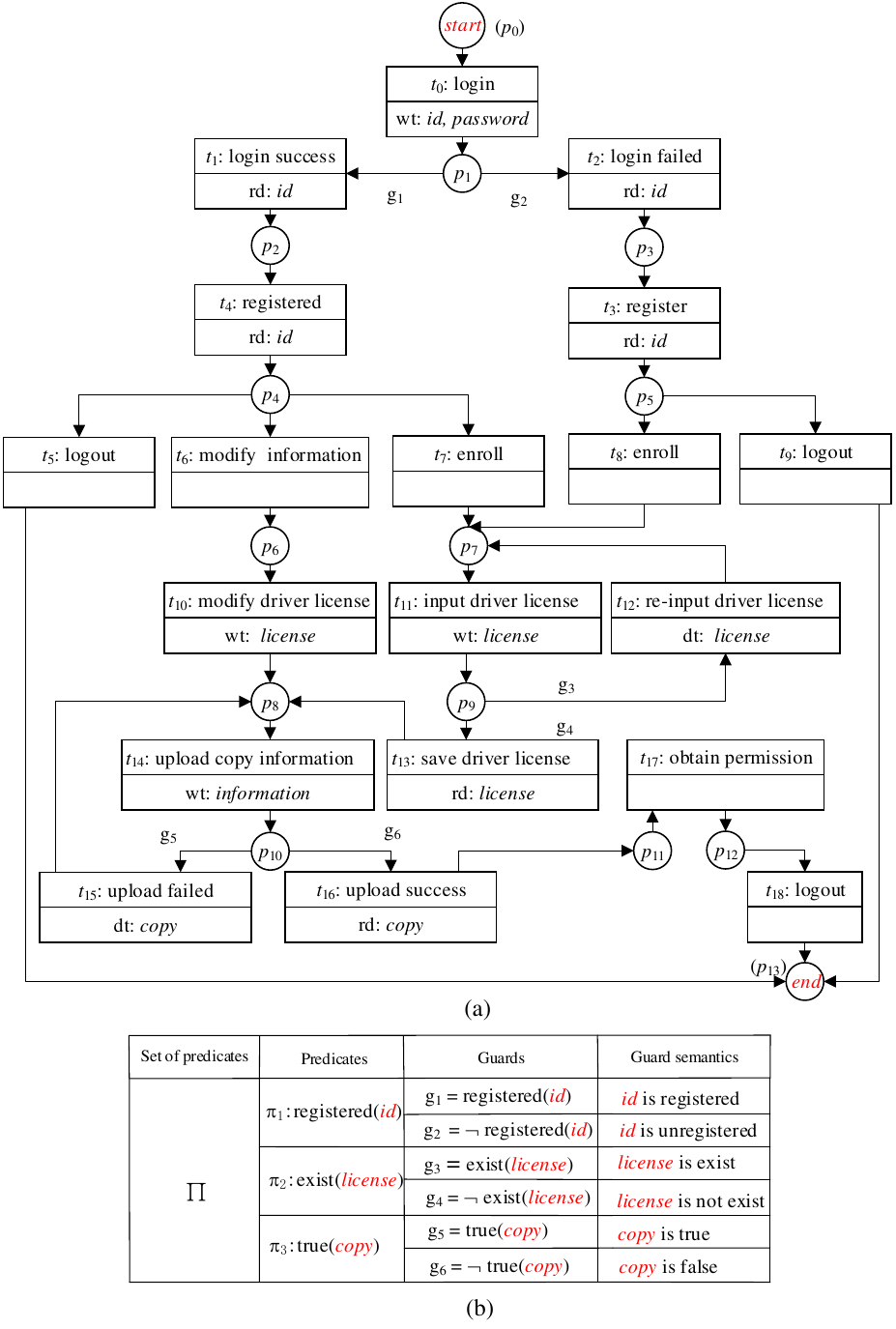}
	\caption{(a) WFD-net modeling Vehicle Management System; (b) Guards.}
	\label{fig02}
	\end{figure}

Since the construction process of an SRG of a WFD-net or a WFT-net do not consider the constraints among different but related guards, it will produce pseudo states. When the following motivation case is modeled by the WFD-net or the WFT-net, its SRG will have the same defect (i.e., pseudo states). Therefore, we use the WFD-net model of this motivation case to illustrate this defect. A university vehicle management system is modeled with the WFD-net in Figure~\ref{fig02}. Places \textbf{\textit{start}} and \textbf{\textit{end}} represent the initial and final state, respectively. When an employee inputs an account number($id$) and a password, s/he can login the system ($t_0$). If the employee has been previously registered, s/he can successfully login the system ($t_1$), open the registration interface ($t_4$), and access the system directly. Otherwise, the login is failed ($t_2$), and the employee must register a new account to access the system ($t_3$). After entering the system, s/he can exit the system ($t_5$ and $t_9$). An employee can need modify her/his existing driver's license information ($t_6$) if s/he has registered these information, or s/he registers her/his driver's license information ($t_7$). Note that the purpose of registering a driver license is to allow her/his car to enter the campus of this university; but these processes are not described here for simplicity. A new user can only choose to register her/his driver's license information ($t_8$). When the employee chooses to modify information, a new driver's license information needs to be re-entered, and the modified information will be saved in the system ($t_{10}$). When the employee registers her/his driver's license information ($t_{11}$), the system will check if this license has been in the system. If yes, the employee is required to re-enter another license information ($t_{12}$), otherwise it will be stored in the system ($t_{13}$). When this checking is approved, the employee needs to submit a copy of id card and some other materials ($t_{14}$). If the materials are incorrect or incomplete after reviewed by a manager, they ($t_{15}$) are required to re-upload. If the materials are approved ($t_{16}$), a permission certification of allowing her/his car enter the school is produced ($t_{17}$). Figure~\ref{fig02} shows the WFD-net model of this workflow system. If we add into the model a table recording the information of users as well as some table operations on the transitions, then we can obtain a WFT-net. Later we will see such a similar net, but here we can illustrate the flaw of the methods of generating its SRG, i.e., generating some pseudo states.

\begin{figure}[tt]
	\centering
	\includegraphics[width=0.36\textwidth]{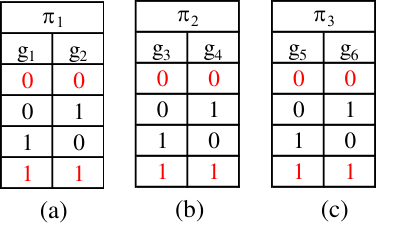}
	\caption{The distribution tables of guards values without constraints.}
	\label{fig03}
\end{figure}

\begin{figure}[tt]
	\centering
	\includegraphics[width=0.48\textwidth]{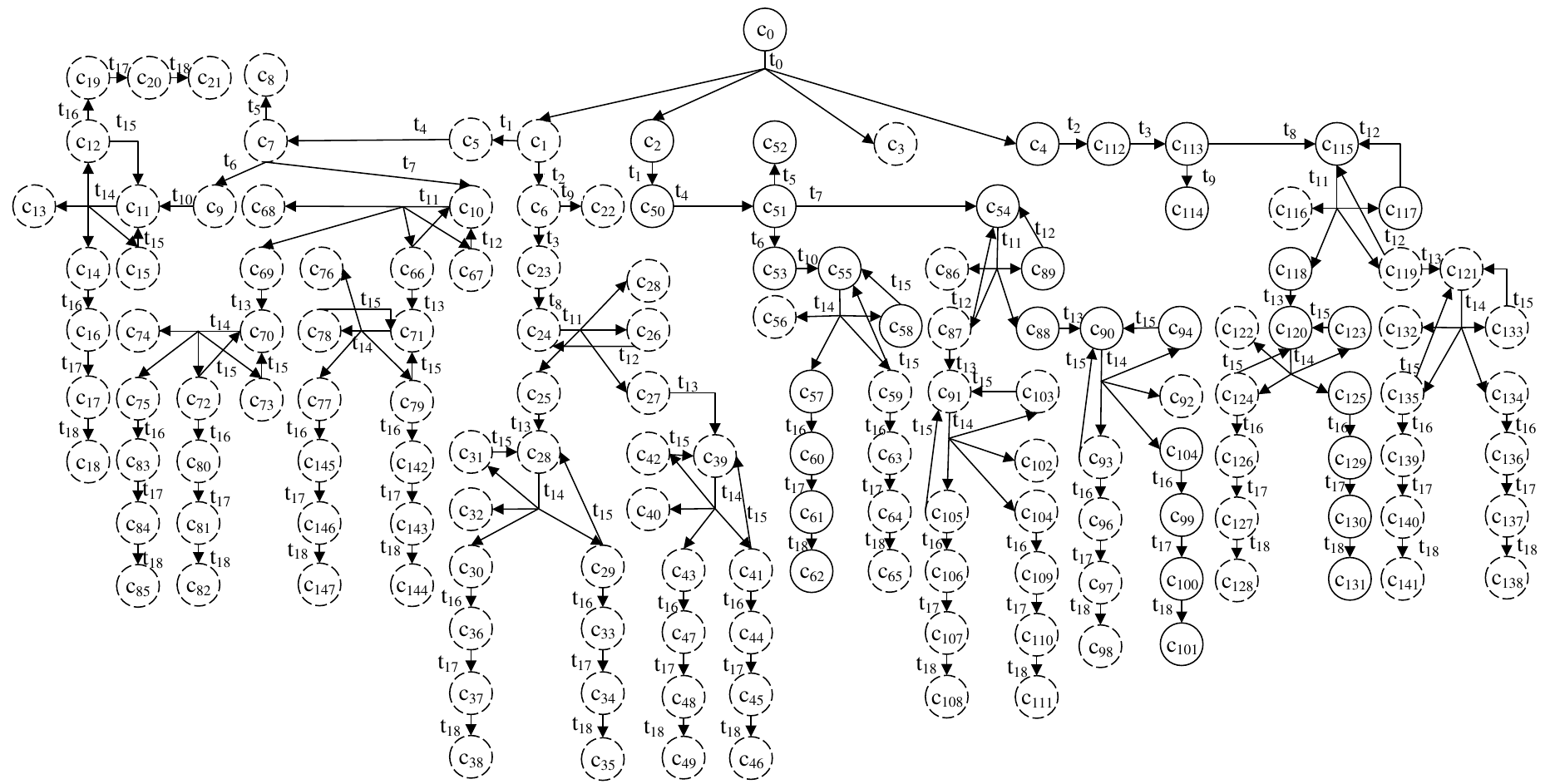}
	\caption{\centering{The state reachability graph of the WFD-net in Fig.\ref{fig02}.}}
	\label{fig04}
\end{figure}

In a WFD-net or a WFT-net, according to their firing rules \cite{ref15,ref23,ref44}, the $read/write$ operations on a transition $t$ is associated with data items in $k$ guards. Since their SRG generation methods do not consider the constraint between these guards, they need to consider all possible assignments values of these guards and thus produce $2^{k}$ states including pseudo states. For example of the $write$ operation on the data item $id$ at $t_0$ in Figure \ref{fig02}~(a), g$_1$ and g$_2$ are associated with $id$. Therefore, when $t_0$ is fired at the initial state $c_0$, guards g$_1$ and g$_2$ have $2^{2}=4$ possible values and thus 4 possible states ($c_1$, $c_2$, $c_3$, and $c_4$ in Figure~\ref{fig04}) are generated. However,  $c_1$ and $c_3$ are pseudo states. Three distribution tables of guards values are generated according to the methods in \cite{ref15,ref23,ref44}, as shown in Figure~\ref{fig03}. The truth table in Figure~\ref{fig03}~(a) describes all possible values of g$_1$ and g$_2$ where $0$ means \textbf{\textit{false}} (short: $F$) and $1$ means \textbf{\textit{true}} (short: $T$). Notice that the red values in Figure~\ref{fig03} should not have occurred. In fact, g$_1$ and g$_2$ in this case are mutually exclusive so that they cannot take the same value at the same time, i.e., neither g$_1=1$ $\wedge$ g$_2=1$ nor g$_1=0$ $\wedge$ g$_2=0$ can take place, but they correspond to $c_1$ and $c_3$ in Figure \ref{fig04}, respectively. Similarly, $\pi_2$ and $\pi_3$ are analyzed, and the results are shown in Figure~\ref{fig03} (b) and (c). As shown in Figure~\ref{fig04}, 147 states are generated, but 113 are pseudo states represented by dotted circles, and the details of each state are shown in Table~\ref{table1}.

According to the definition of soundness of a WFD-net, the final state \textbf{\textit{end}} is always reachable from the initial state \textbf{\textit{start}} in Figure~\ref{fig02} (a) \cite{ref15,ref16,ref17,ref35}. However, the control flow of this workflow has a flaw actually but it cannot be reflected by soundness. For example, when a user $id1$ register $license1$, the system finds that $license1$ has been registered by others (we assume that this license has been registered by the employee with $id2$). Therefore, the $id1$ needs to register a new license. S/He may register with the information of $license2$ first and then successfully modifies $license2$ into $license1$. This case is possible since the modification process of this system has not the step of checking if a modified license has been registered by others. This is also the reason why \cite{ref39} defines WFT-net. In other words, to find such a control-flow flaw, a model must consider the changes of data items of some users. 

Literature \cite{ref39} has been able to detect this control flow defect, because it only defines soundness, some design requirements related to data items cannot be expressed and detected, e.g., $AG(\forall r \in R, r.id\neq \emptyset\rightarrow r.license \neq \emptyset)$. Such design requirements can be expressed by our database-oriented  computation tree logic (DCTL) and detected using our model checking method. The analysis of the motivation case reveals two problems with existing workflow models: (1) their analysis methods can generate pseudo states; (2) these methods cannot represent some data-item-related design requirements.

\section{WFTC-net and SRG}
A WFT-net has the same defect as WFD-net, i.e., its SRG has pseudo states. If there are unreasonable states in the SRG, then the correctness of model verification results cannot be guaranteed. Therefore, in order to avoid pseudo states, we propose a new model called workflow net with tables and constraints and a new refinement method to generate the accurate SRG. This method effectively removes pseudo states and enables more accurate description of system behaviors.

\subsection{Workflow Net with Tables and Constraints}
WFT-net and WFD-net do not consider the constraints between guards, and they produce pseudo states. In order to solve this problem, WFT-net is re-defined and these constraints are considered in this new definition.

\begin{definition}[Constraint Sets]
	Given a set of predicates $\prod=\{\pi_1, \pi_2,...,\pi_n\}$, a predicate $\pi_i$ may be associated with multiple guards. When different guards perform data operations on the same data items, it is necessary to consider the constraint relationship between guards and form a proposition formula $\omega$ of guards through operators $\wedge$, $\vee$, and $\neg$. The $\vee$ operator combines these propositional formulas to form a constraint $true$, and $\textbf{\textit{Res}}$ is a set of constraints.
	\label{defn4}
\end{definition}

\begin{figure} [tt]
	\centering
	\includegraphics[width=0.36\textwidth]{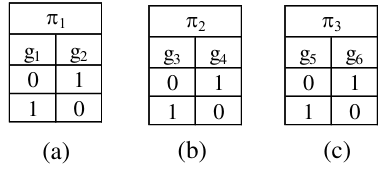}
	\caption{The distribution table of guards values with constraints.}
	\label{fig05}
\end{figure}

\begin{figure}[tt]
	\centering
	\includegraphics[width=0.48\textwidth]{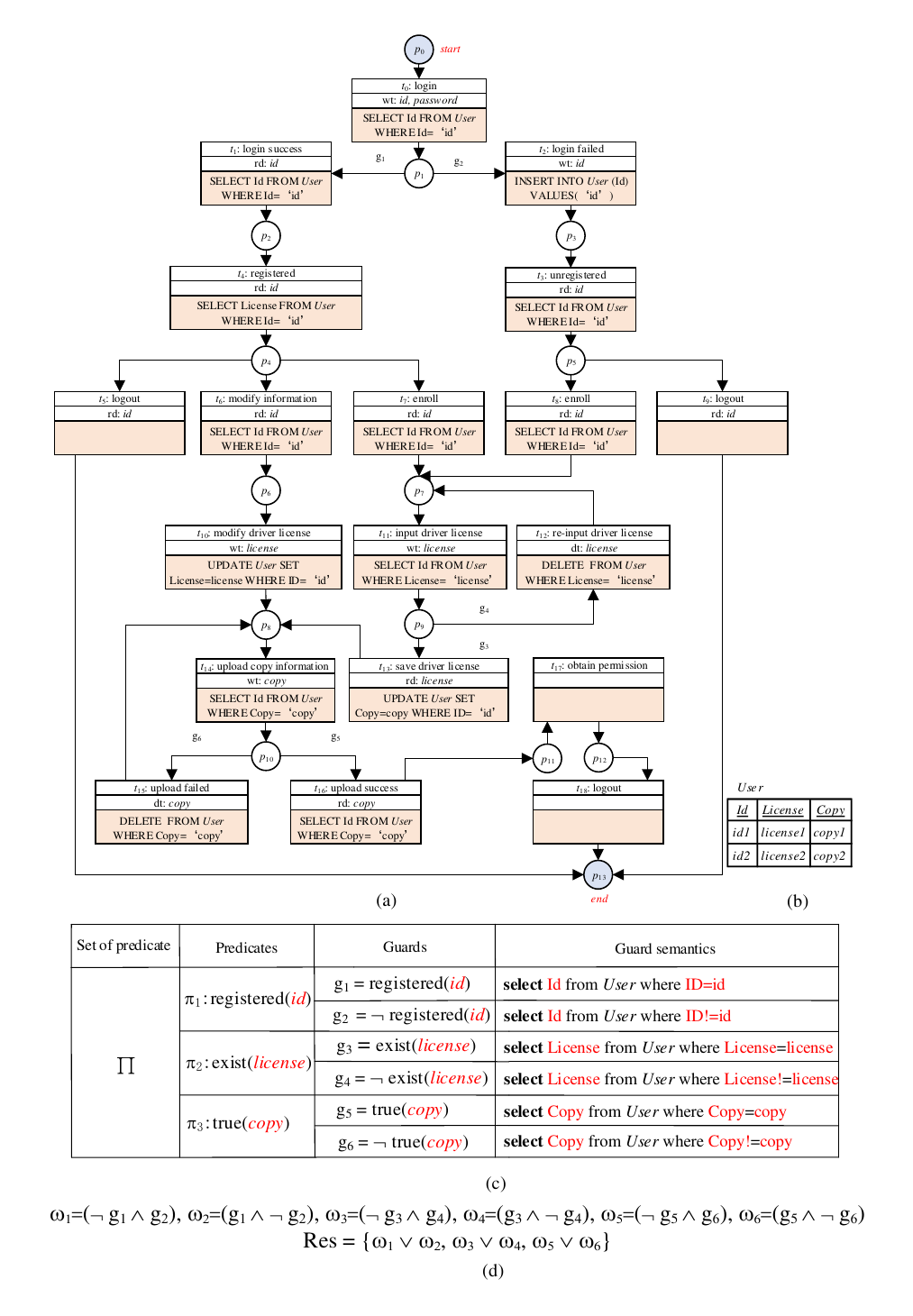}
	\caption{(a) A WFTC-net; (b) An initial table; (c) Guards; (d) Constraint set.}
	\label{fig06}
\end{figure}

Constraint set $\textbf{\textit{Res}}$ can be used in any model. As shown in Figure~\ref{fig02} (b) or Figure~\ref{fig06} (c), there is a set of predicates $\prod=\{\pi_1, \pi_2, \pi_3\}$. Since g$_1$ and g$_2$ are associated with predicate $\pi_1$ and are both about the $write$ operation on the data item $id$, there exists a constraint between them. Therefore, we define two proposition formulas, i.e., $\omega_1$ = $ $g$_1 \wedge \neg$ g$_2$ and $\omega_2$ = $\neg$ g$_1\wedge$ g$_2$. $\omega_1$ means that if g$_1$ is $true$, then g$_2$ is not $true$, and $\omega_2$ means that if g$_2$ is $true$, then g$_1$ is not $true$. Therefore, the constraint $true = \omega_1 \vee \omega_2$ is constructed, which means that g$_1$ and g$_2$ cannot have the identical value at any time. According to constraint we know that $(1,1)$ and $(0,0)$ in Figure~\ref{fig03} (a) can not take place, i.e., $c_1$ and $c_3$ in Figure~\ref{fig04} are pseudo states and should be deleted. Similarly, we should have the following proposition formulas: $\omega_3$ = g$_3$ $\wedge$ $\neg$ g$_4$, $\omega_4$ = $\neg$ g$_3$ $\wedge$ g$_4$, $\omega_5$ =  g$_5$ $\wedge$ $\neg$ g$_6$, and $\omega_6$ = $\neg$ g$_5$ $\wedge$ g$_6$, and thus construct the following two constraints $true= \omega_3 \vee \omega_4$ and $true= \omega_5 \vee \omega_6$. Figure~\ref{fig05} shows the values of guards under these constraints. The constraint set $\textbf{\textit{Res}}=\{\omega_1 \vee \omega_2, \omega_3 \vee \omega_4, \omega_5 \vee \omega_6\}$ is constructed. After removing the dotted circles out of Figure~\ref{fig04} according to these constraints, a correct SRG is obtained.

\begin{definition}[Workflow Net with Tables and Constraints]
	A workflow net with tables and constraints (WFTC-net) is a 2-tuple $N = (N'$, $\textbf{\textit{Res}})$=  
	$(P$, $T$, $F$, $G$, $D$, $R$, $rd$, $wt$, $dt$, $sel$, $ins$, $del$, $upd$, $guard$, $\textbf{\textit{Res}})$ where
	\begin{itemize}
		\item[(1)] $N'$ is a WFT-net; and
		\item[(2)] $\textbf{\textit{Res}}$ is a set of constraints.
	\end{itemize}	
	\label{defn6}
\end{definition}

For example, Figure~\ref{fig06} is a WFTC-net that considers table operations and guard constraints on the basis of the WFD-net in Figure~\ref{fig02}. Figure~\ref{fig06} (a) describes the business logic of our motivation example, including the operations of its data items and table, and guards associated with transitions; (b) shows an initial table; (c) shows guards related to data and table operations; and (d) lists the constraints satisfied by these guards. In this WFTC-net:

(1) $D=\{id$, $password$, $license$, $copy\}$ is a set of data items, which can be viewed as intermediate variables via which a user interacts with this system;

(2) $R=User=\{r_1$, $r_2\}$ is an initial table, where each record is composed of three attributes, i.e., $Id$, $License$, and $Copy$;

(3) A $write$ operation on the data item $id$ is associated with the transition $t_0$, i.e., $wt(t_0)=\{id, password\}$, and a $select$ operation on the attribute $Id$ in $User$ is also associated with $t_0$, i.e., $sel(t_0)=\{id1, id2\}$. Obviously, $upd(t_0)=\{\varnothing\}$, $del(t_0)=\{\varnothing\}$, and $ins(t_0)=\{\varnothing\}$; and

(4) $\prod=\{\pi_1$, $\pi_2$, $\pi_3\}$ is a set of predicates and the guard set is $G_{\pi}$ = $\{$g$_1$, g$_2$, g$_3$, g$_4$, g$_5$, g$_6\}$. We define proposition formulas $\omega_1$ = $\neg$ g$_1$ $\wedge$ g$_2$, $\omega_2$ = g$_1$ $\wedge$ $\neg$ g$_2$, $\omega_3$ = $\neg$ g$_3$ $\wedge$ g$_4$, $\omega_4$ = g$_3$ $\wedge$ $\neg$ g$_4$, $\omega_5$ = $\neg$ g$_5$ $\wedge$ g$_6$, $\omega_6$ = g$_5$ $\wedge$ $\neg$ g$_6$, and thus the constraint set is $\textbf{\textit{Res}} = \{\omega_1 \vee \omega_2$, $\omega_3 \vee \omega_4$, $\omega_5 \vee \omega_6 \}$. Notice that the meaning of these predicates and guards are listed in Figure~\ref{fig06} (d) and we always require that the value of a constraint is $true$. 

A state is a smallest unit of an SRG, reflecting the detailed information in each state. A state in WFTC-net consists of a marking $m$, abstract data operation, data items in a table, and guards.

\begin{definition}[State]
	Let $N = (P$, $T$, $F$, $G$, $D$, $R$, $rd$, $wt$, $dt$, $sel$, $ins$, $del$, $upd$, $guard$, $\textbf{\textit{Res}})$ be a WFTC-net, $c=\langle m$, $\theta_D$, $\vartheta_R$, $\sigma\rangle$ is a state of $N$, where
	\begin{itemize}
		\item[(1)] $m$ is a marking of $N$;
		\item[(2)] $\theta_D$: $D \rightarrow \{ \bot$, $\top \}$ is the values of the data items in the current state. At the initial state, every data item is \emph{\textbf{\textit{undefined}}} which is represented as $\bot$. A write operation is executed on a data item, it is assigned a value which means that it is \emph{\textbf{\textit{defined}}} that is represented by $\top$;
		\item[(3)] $\vartheta_R$: $R \rightarrow \left \{ \bot,\top \right \}$ is the values of the data items in the current table situation. Notice that if an attribute of a record in the current table is not assigned a value, it is \emph{\textbf{\textit{undefined}}} $(\bot)$. An update or an insert operation is executed on an attribute of a record in the current table is assigned a concrete value, it is \emph{\textbf{\textit{defined}}} $(\top)$. In a WFTC-net, a table is usually defined, and each data item in the table is a defined concrete value; and
		\item[(4)]  $\sigma:\Pi \rightarrow \left \{ true, false, \bot\right\}$ represents the assignment of each predicate. Since each predicate is associated with some data items, after all related data items of a predicate are written, this predicate is \textbf{\textit{true}} ($T$) or \textbf{\textit{false}} ($F$). Otherwise, its value is \emph{\textbf{\textit{undefined}}} $(\bot)$. According to the values of these predicates, we can compute the value of each guard under the given constraints. In a state, we list the values of all guards rather than the values of all predicates.	
	\end{itemize}
	\label{defn7}
\end{definition}

For a WFTC-net in Figure~\ref{fig06}, its initial state is $c_0=\langle \emph{\textbf{\textit{start}}}$, $\{id=\bot$, $password=\bot$, $license=\bot$, $copy=\bot\}$, $\{(id1$, $license1$, $copy1)$, $(id2$, $license2$, $copy2)\}$, $\{$g$_1$ = $\bot$, g$_2$ = $\bot$, g$_3$ = $\bot$, g$_4$ = $\bot$, g$_5$ = $\bot$, g$_6$ = $\bot\}\rangle$. In this initial state, only a place \textit{\textbf{start}} contains a token, where $\theta_D$ assigns $\bot$ to all data items $id$, $password$, $license$, and $copy$, i.e., all data items are undefined. Additionally, the table situation is the same as that in Figure \ref{fig06} (b), which means that the values of the data items in $\vartheta_R$ are defined and all guards are undefined.

\subsection{Generating the SRG of a WFTC-net using a Data Refinement Method}
The main reason of generating pseudo states in existing methods are that constraints of guards are not considered in the procedure of generating new states. Therefore, this section provides a new data refinement method to avoid this situation.

\begin{definition}[Data Refinement]
	Let $N = (P$, $T$, $F$, $G$, $D$, $R$, $rd$, $wt$, $dt$, $sel$, $ins$, $del$, $upd$, $guard$, $\textbf{\textit{Res}})$ be a WFTC-net, $c=\langle m$, $\theta_D$, $\vartheta_R$, $\sigma\rangle$ be a state and $d \in D$ be a data item. When $t$ is enabled at $c$, we use $REF(c$, $d)$ to represent a refinement of $d$ at $c$. Given a table $R$ and an attribute $\underline{d}$ of it, $R(\underline{d})$ denotes the set of values occuring in $R$ and corresponding to $\underline{d}$. If there exists a data item $d$ at $t$ needs to be modified via some operation: $d \xrightarrow[t]{\textbf{mo}} d_{new}$, where $\textbf{mo}$ $\in$ $\{wt,ins,del,upd\}$, such that if $d_{new}$ $\in$ $R$, then it doesn't need to be refined, $REF(c$, $d)$ = $R($\underline{d}$)$. Otherwise, it needs to be refined, $REF(c$, $d)$ = $R($\underline{d}$)$ $\cup$ $\{d_{new}\}$ where $d_{new}$ is a new value not occurring in R(\underline{d}).
	\label{defn7.1}
\end{definition}

The initial state of a WFTC-net is denoted as $c_0=\langle \emph{\textbf{\textit{start}}}$, $\{id=\bot$,  $password=\bot$, $license=\bot$, $copy=\bot\}$, $\{(id1$, $license1$, $copy1)$, $(id2$, $license2$, $copy2)\}$, $\{$g$_1$ = $\bot$, g$_2$ = $\bot$, g$_3$ = $\bot$, g$_4$ = $\bot$, g$_5$ = $\bot$, g$_6$ = $\bot\}\rangle$ in Figure~\ref{fig06}. According to Definition \ref{defn7.1}, the transition $t_0$ is enabled at the initial state $c_0$. After firing $t_0$, a data item $id$ is written at $t_0$, i.e. $id \xrightarrow[t_0]{wt} id_{new}$, so it should be refined. Because $R(\underline{id})=\{id1, id2\}$, when the $id$ = $id1$ or $id2$, we know that $id1$ $\in$ $R(\underline{id})$ or $id2$ $\in$ $R(\underline{id})$, it doesn't need to be refined, i.e. $REF(c_0, id)$ = $R(\underline{id})$ = $\{id1, id2\}$. However, when the $id$ = $id3$, i.e. register a new user, since $id3$ $\notin$ $R(\underline{id})$, it needs to be refined, we have $REF(c_0, id)$ = $R(\underline{id}) \cup \{id3\}$ = $\{id1, id2, id3\}$. This means that the person using this system is possibly an existing user ($id1$ or $id2$) or a non-exiting one (we name it as $id3$).     

After firing a transition $t$ at state $c$, if a refined data item $d$ is related to a table $R$, then $\vartheta_R$ in $c$ will be also changed accordingly. For example, performing a $select$ operation on $id$ at $t_0$ will not change the values of all attributes in the $R$. Therefore, the record $(id1,license1,copy1)$ and $(id2,license2,copy2)$ in the table remain unchanged. A new tuple $\{id3$, $\bot$, $\bot\}$ will be inserted in the $R$ since a new data item $id3$ is inserted at $t_2$. At the same time, the $R$ will be updated with a new $R'$ = $\{(id1$, $license1$, $copy1)$, $(id2$, $license2$, $copy2)$, $(id3$, $\perp$, $\perp)\}$. For convenience, we define $ins(R')$ to perform an $insert$ operation on an element in the $R'$, $del(R')$ to perform a $delete$ operation on an element in the $R'$, $sel(R')$ to perform a $select$ operation on an element in the $R'$, and $upd(R')$ to perform an $update$ operation on an element in the $R'$. g$_d$ represents the operation performed by guard $g$ on data item $d$. We are inspired by \cite{ref15,ref39,ref42} and Definition \ref{defn7.1}, the firing rules of WFTC-net is defined as follows.

\begin{definition}[The firing rule of WFTC-net]
	Let $N = (P$, $T$, $F$, $G$, $D$, $R$, $rd$, $wt$, $dt$, $sel$, $ins$, $del$, $upd$, $guard$, $\textbf{\textit{Res}})$ be a WFTC-net. A transition $t\in T$ is enabled at a state $c=\langle m$, $\theta_D$, $\vartheta_R$, $\sigma\rangle$, which is denoted by $c[t\rangle$, if and only if:
	\begin{itemize}
		\item[(1)] $\forall$ $t$ $\in$ $T:$ $m[t\rangle;$
		\item[(2)] 
		$\forall$ $d\in$ $(rd(t)$ $\cap$ $D):$ $\theta_D$$(d)$ = $\top;$ 
		$\forall$ $d\in$ $(wt(t)$ $\cap$ $D):$ $\theta_D$$(d)$ = $\top;$ 
		$\forall$ $d\in$ $(dt(t)$ $\cap$ $D):$ $\theta_D$$(d)$ = $\bot;$
		\item[(3)] 
		$\forall$ $d\in$ $(R$ $\cap$ $del(t)):$ $\vartheta_R(d)=\bot;$ 
		$\forall$ $d\in$ $(R$ $\cap$ $(sel(t)$ $\cup$ $ins(t)$ $\cup$ $upd(t))):$ $\vartheta_R(d)=\top;$ and
		\item[(4)] 
		$\sigma(G(t))=true.$
	\end{itemize}

	After firing a transition t, the $t$ is enabled at state $c$. A new state  $c'=\langle m'$, $\theta_D'$, $\vartheta_R'$, $\sigma'\rangle$ is generated, which is denoted as $c[t\rangle c'$, such as:
	\begin{itemize}
		\item[(1)] $m[t\rangle m';$
		\item[(2)] $\forall$ $d\in$ $dt(t):$ $\theta_D'(d)$ = $\bot;$ $\forall$ $d\in$ $(wt(t)\cup rd(t)):$ $\theta_D'(d)$ = $\top;$ $\forall$ $d'\in$ $REF(c, d):$ $\theta_D'(d)$ = $\theta_D(d)$ $\cup$ $\{d'\};$
		\item[(3)] $\forall$ $d\in$ $(wt(t)\setminus dt(t)):$ $\theta_D'(d)=\top;$ $\forall$ $d\in$ $((D$ $\cup$ $REF(c, d))$ $\setminus$ $(dt(t)$ $\cup$ $wt(t))):$ $\theta_D'(d)=\theta_D(d);$
		\item[(4)] $\forall$ $R' \in ins(t),$ $\forall$ $ins(R') \cap R \ne \{\varnothing\}:\vartheta_R'(R') = \vartheta_R (R);$ $\forall$ $R' \in ins(t)$, $\forall$ $ins(R') \cap R=\emptyset:\vartheta_R'(R') = \vartheta_R (R) \cup ins(R');$
		\item[(5)] $\forall$ $R'\in$ $del(t),$ $\forall$ $del(R') \subseteq R:\vartheta_R '(R') = \vartheta_R (R)\setminus del(R');$ $\forall$ $R'\in del(t),$ $\forall$ $del(R')\nsubseteq R:\vartheta_R '(R') = \vartheta_R (R);$	
		\item[(6)] $\forall$ $R'\in upd(t),$ $upd(R')$ $\subseteq$ $R:$ $\vartheta_R '(R')$ = $(\vartheta (R)\setminus upd(R')) \cup upd(R');$		
		\item[(7)] $\forall$ $R' \in sel(t),$ $\forall$ $sel(R') \subseteq R:\vartheta_R '(R') = \top$; $\forall$ $R' \in sel(t),$ $\forall$ $sel(R')\nsubseteq R:\vartheta_R '(R')=\bot ;$
		\item[(8)] $\forall$ g$_d \in$ $G,$ $d \in$ $(wt(t)$ $\cup$ $rd(t)$ $\cup$ $upd(t)$ $\cup$ $ins(t)$ $\cup$ $sel(t)$ $\cup$ $dt(t)$ $\cup$ $del(t)$ $\cup$ $REF(c,d)):$ $\sigma'$$($g$_d)$ = $\{true, false\};$
		\item[(9)] $\forall$ g$_d \in G,$ $d \in$ $REF(c,d)$ $\wedge$ g$_d$ $\in$ $\textbf{\textit{Res}}:$ $\sigma'$$($g$_d$$)$ = $\{true\};$ $\forall$ g$_d \in G,$ $d \in$ $REF(c,d)$ $\wedge$ g$_d$ $\notin$ $\textbf{\textit{Res}}:$ $\sigma'$$($g$_d$$)$ = $\{false\};$ 
		\item[(10)] $\forall$ g$_d \in G,$ $d \in (\theta_D \cup \vartheta_R)$ $\wedge$ $g_d $ = $ \bot:$ $\sigma'$$($g$_d$$)$ = $\{\bot\};$ and
		\item[(11)] $\forall$ g$_d \in G,$ $((\theta_D \cup \vartheta_R)\cap REF(c,d))$ = $\{\varnothing\}:$ $\sigma'$$($g$_d$$)$ = $\sigma$$($g$_d$)$.$ 
	\end{itemize}	
	\label{defn8}
\end{definition}

For example, given an initial state $c_0$ = $\langle \emph{\textbf{\textit{start}}}$, $\{id=\bot$, $password=\bot$, $license=\bot$, $copy=\bot\}$, $\{(id1$, $license1$, $copy1)$, $(id2$, $license2$, $copy2)\}$, $\{$g$_1$ = $\bot$, g$_2$ = $\bot$, g$_3$ = $\bot$, g$_4$ = $\bot$, g$_5=\bot$, g$_6$ = $\bot\}\rangle$ of the WFTC-net in Figure~\ref{fig06} (a), the transition $t_0$ is enabled at $c_0$. After firing $t_0$ at $c_0$, the token in $\textbf{\textit{start}}$ will be moved into $p_1$. Data items $id$ and $password$ are written at $t_0$. According to Definition \ref{defn7.1}, because $id$ belongs to the table, it needs to be refined. $id$ is defined and $REF(c, id)=\{id1,id2,id3\}$ is obtained. There is no $read$ or $write$ operation performed on data elements $license$ and $copy$, so their values remain unchanged, i.e., $license=\bot$ and $copy=\bot$. Since the data items in the table are not modified at $t_0$, they remain the same, i.e., $User$ = $\{(id1$, $license1$, $copy1)$, $(id2$, $license2$, $copy2)\}$. Guards g$_1$ and g$_2$ are related to the operation of the $id$ at $t_0$, and thus we can calculate two proposition formulas, i.e., $\omega_1$ = g$_1$ $\wedge$ $\neg$ g$_2$ and $\omega_2$ = $\neg$ g$_1$ $\wedge$ g$_2$ according to the constraint $true = \omega_1 \vee \omega_2$. Since g$_1$ is associated with the data items in the $User$, it satisfies constraint $true$. In consequence, when $id1$ or $id2$ is written at the transition $t_0$, i.e., $\theta_D(id)=id1$ or $\theta_D(id)=id2$, it makes g$_1=T$ and g$_2$ = $F$. Similarly, for the case of $\theta_D(id)=id3$, we have g$_1$ = $F$ and g$_2$ = $T$, since $id3$ does not belong to the $User$. Because guards g$_3$, g$_4$, g$_5$, and g$_6$ are independent of the $id$, their guard values are still undefined, i.e., g$_3$ = $\bot$, g$_4$ = $\bot$, g$_5$ = $\bot$, and g$_6$ = $\bot$. Therefore, firing $t_0$ generates three new configurations:\\
$c_1$ = $\langle \emph{\textbf{\textit{start}}}$, $\{id1$, $password$, $license=\bot$, $copy=\bot\}$, $\{(id1$, $license1$, $copy1)$, $(id2$, $license2$, $copy2)\}$, $\{T$, $F$, $\bot$, $\bot$, $\bot$, $\bot\}\rangle$, \\
$c_2$ = $\langle \emph{\textbf{\textit{start}}}$, $\{id2$, $password$, $license=\bot$, $copy=\bot\}$, $\{(id1$, $license1$, $copy1)$, $(id2$, $license2$, $copy2)\}$, $\{T$, $F$, $\bot$, $\bot$, $\bot$, $\bot\}\rangle$,~and\\
$c_3$ = $\langle \emph{\textbf{\textit{start}}}$, $\{id3$, $password$, $license=\bot$, $copy=\bot\}$, $\{(id1$, $license1$, $copy1)$, $(id2$, $license2$, $copy2)\}$, $\{F$, $T$, $\bot$, $\bot$, $\bot$, $\bot\}\rangle$.

\begin{figure}[tt]
	\centering
	\includegraphics[width=0.40\textwidth]{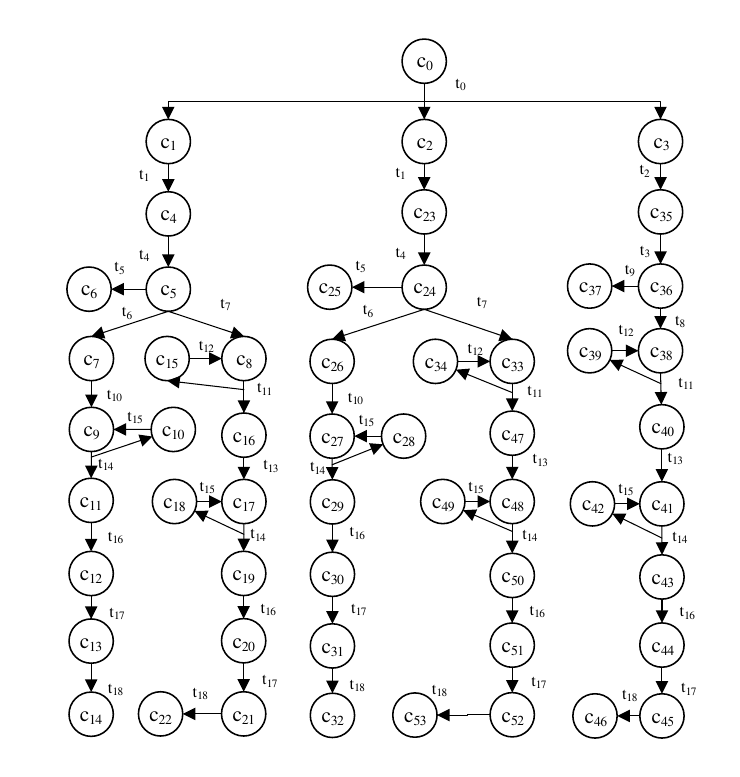}
	\centering
	\caption{State reachability graph of WFTC-net.}
	\label{fig07}
\end{figure}

Subsequently, transition $t_2$ is enabled at $c_3$. After firing $t_2$ , the token in $p_1$ is moved into $p_3$, and the new user $id3$ is inserted in $R$. Therefore, the data elements in the table are updated from $R=\{(id1$, $password$, $license1$, $copy1)$, $(id2$, $license2$, $copy2)\}$ to $R'=\{(id1$, $license1$, $copy1)$, $(id2$, $license2$, $copy2)$, $(id3$, $\bot$, $\bot)\}$. By Definition~\ref{defn8}, the values of guards in the new state only need to be consistent with that in $c_3$, since $t_2$ has no guards. Through the above analysis, a new state $c_{35}=\langle p_3$, $\{id3$, $password$, $license=\bot$, $copy=\bot\}$, $\{(id1$, $license1$, $copy1)$, $(id2$, $license2$, $copy2)$, $(id3$, $\bot$, $\bot)\}$, $\{F$, $T$, $\bot$, $\bot$, $\bot$, $\bot\}\rangle$ is generated. 
Similarly, we get new states $c_4$ and $c_{23}$ by firing $t_1$ at states $c_1$ and $c_{23}$. As shown in Figure~\ref{fig07}, 54 states are generated and the details of each  state are shown in Table~\ref{table2}.

For a WFTC-net $N$, $R(N, c_0)$ is all reachable states of $N$ that forms the node set of the SRG where $c_0$ is called the root node. $E$ is the set of all directed edges that are labeled by transitions. According to transition enabling condition and firing rule, we propose an algorithm for generating the SRG of the $N$, as shown in Algorithm~\ref{alg:2}.

\begin{algorithm}[tt]
	\caption{A Novel Data Refinement Method to Generate SRG}
	\hspace*{0.02in} {\bf Input:} WFTC-net $N$ \\
	\hspace*{0.02in} {\bf Output:} SRG
	\begin{algorithmic}[1]
		\State $c_{new}$ $\leftarrow$ $\emptyset$, $R(N, c_0)$ $\leftarrow$ $\emptyset$, $E$ $\leftarrow$ $\emptyset$;
		\State $\textbf{\textit{PutTail}}$$(c_{new}, c_0)$;	
		\While {$c_{new}$ $\neq$ $\emptyset$}
		\State $c$ $\leftarrow$ $\textbf{\textit{GetHead}}$$(c_{new})$; 
		\State $R(N, c_0)$ $\leftarrow$ $R(N, c_0)$ $\cup$ $\{c\}$;
		\For{each $t$ $\in$ $T$ such that $c[t$$\rangle$}
		\State Produce marking $c'$ such that $c[t$$\rangle$c' according to Definition~\ref{defn8};
		\If{c' $\notin$ $c_{new}$ $\cup$ $R(N, c_0)$}
		\State $\textbf{\textit{PutTail}}$$(c_{new},$ $c')$;
		\EndIf
		\State  $E$ $\leftarrow$ $E$ $\cup$ $\{(c, c')_t\}$;
		\EndFor
		\EndWhile			
	\end{algorithmic}
\label{alg:2}
\end{algorithm}

Based on Algorithm~\ref{alg:2}, an SRG is shown in Figure~\ref{fig07} where the details of each state are shown in Table~\ref{table2}. The SRG generated based on the guard constraints can restrict the state space to a certain extent and effectively remove pseudo states, which is more in line with actual demands when characterizing the system's dynamic behavior. The validation of the correctness of a workflow model will be discussed in the next section.

\section{Logical verification based on WFTC-net state reachability graph}

We know that the correctness of a workflow model is closely related to the data requirements \cite{ref32}. Since a motivation case lacks the verification of data operation and requirements, it contains logical defects, so that the correctness of the model cannot be guaranteed. During system operation, user operations cause changes to the data in the background database. When we model the system related to database tables, some logical defects and design specifications in the system cannot be detected by traditional methods. In order to further verify the logic errors and improve the correctness of the workflow model. This section provides a database-oriented computation tree logic (DCTL) method, which can effectively verify the logical errors in the model.

\subsection{Syntax and Semantics of DCTL}

In a real system, the design specifications are related to the data operations of multiple process instances. The system is usually abstracted from the corresponding Kripke structure or PN, and the specifications are represented by temporal logic, e.g., CTL, LTL. By exhaustively searching each state in the state space, results are given that confirm whether the system model satisfies the system specification \cite{ref44,ref45,ref51}. A vehicle management system, as shown in Figure~\ref{fig06}, to verify a workflow system related to database tables operations, such as specification, i.e., $\varphi_1=AG(\forall(id1\in R, id2\in R), [id1\neq id2 \rightarrow id1.license1\neq id2.license2])$. Based on the above analysis, we know that when the tables are in dynamic change, the tables are not deterministic and cannot be described concretely. Traditional CTL formulas cannot describe requirements related to database operations. To solve such problems, we extend the traditional CTL and named it DCTL. The DCTL retains the syntax elegance of the original CTL and significantly enhances its expression ability.

The syntax of DCTL is defined based on WFTC-net. Secondly, the syntax, semantics, and truth values of DCTL formulas are explained and verified on its SRG. The finite SRG structure is a 5-tuple SRG = $\langle AP, C, \Re, c_0, L \rangle$, where:

(1) $AP$ is a set of atomic propositions;

(2) $C$ is a finite set of states;

(3) $\Re\subseteq C\times C$ is a transition relation, and it must be complete, i.e., for all states $c\in C$, there is a state $c'\in C$ that satisfies $(c,c')\in\Re$;

(4) $c_0\in C$ is an initial state; and

(5) $L:C\rightarrow 2^{AP}$ is a labeling function used to label the set of atomic propositions satisfied on a state. 

An SRG of WFTC-net as shown in Figure~\ref{fig07}, a set of atomic propositions $AP$ = \{$p_0$, $...$, $p_{13}$, $id1$, $id2$, $license1$, $license2$, $copy1$, $copy2$\}, a set of states $C$ = $\{c_0$, $c_1$, $...$, $c_{53}\}$, a state transition relation is $\Re$ = $\{(c_0,c_1)$, $(c_0,c_2)$, $(c_0,c_3)$, $...$, $\}$, and a labeling function $L(c_0)$ = $\{p_0,\{(id1,license1,copy1),(id2,license2,copy2)\}\}$. Other labeling function may be understood similarly.

A path in an SRG is an infinitely long state sequence $\Psi =c_0,c_1,...$, such that for all $i\geq 0$, $c_i \in C$ and $(c_i,c_{i+1})\in R$, $c_0$ is called the initial state of $\Psi$. We can think of path $\Psi$ as an infinite branch of the corresponding computation tree in the SRG. Since the SRG has a finite set of states, one or more states will appear multiple times on a path. In other words, a path represents an infinite traversal over the state transition graph.

A traditional CTL formula is composed of path quantifiers and temporal modalities operators. A path quantifier describes the branching logic of the computation tree. There are two kinds of path quantifiers: $\forall$ (universal path quantifier) and $\exists$ (existential path quantifier). They are used to represent all paths from a state and the properties of some paths, respectively. There are four temporal modalities operators, and they have the following meanings: (1) $X$ $\varphi _1$ $(next)$: Indicates that the next state of the current state satisfies $\varphi _1$; (2) $F$ $\varphi _1$ $(eventually)$: Indicates that a state on the path in the future satisfies $\varphi _1$; (3) $G$ $\varphi _1$ $(always)$: Indicates that all states on the path satisfy $\varphi _1$; ~and $\varphi _1$ $U$ $\varphi _2$ $(until)$: Indicates that there exists a state on the path satisfying $\varphi _2$, and all states before this state satisfy $\varphi _1$.

DCTL is an extension of CTL, and its formal grammar is shown below:

\begin{definition}[Syntax of DCTL]
	Giving a WFTC-net $N = (P$, $T$, $F$, $G$, $D$, $R$, $rd$, $wt$, $dt$, $sel$, $ins$, $del$, $upd$, $guard$, $\textbf{Res\textit{}})$, the DCTL syntax is defined as follows:
	$\Phi::$=$d|$$D_{i}^{j}|$$\exists d$$\in$$\mathbb{R}$,$[\Phi\triangle d]|$$\forall$$ d\in$$\mathbb{R}$,$[\Phi\triangle$$ d]$$|\exists$$ D_{i}^{j} $$\in R$,$[\Phi $$ \triangle D_{i}^{j}]|$$\forall D_{i}^{j} \in R$,$[\Phi \triangle D_{i}^{j}]|\neg \Phi|\Phi\wedge \Phi|\Phi\vee \Phi$; and \\
	$\varphi::$=$true|p|$$\neg\varphi|$$\varphi\wedge\varphi|$$\varphi\vee\varphi|$$\Phi|$$AX\varphi|$$EX\varphi|$$AG\varphi|$$EG\varphi|$ $A(\varphi U \varphi)|$$E(\varphi U \varphi)$, i.e., $p\in P$, $d\in \mathbb{R}$ is natural numbers, $\triangle  \in \{  \le , < , = , \ne , \ge , > \}$ , $\Phi$ and $\varphi$ are represent DCTL and CTL subformulas, respectively.
	\label{defn9}
\end{definition}

$R\in Table$ is a record set containing $n$ tuples and each tuple is m-dimensional. $D_{i}^{j}$ represents the data item of the $j$-$th$ $(1\leq j \leq n)$ dimensional vector of the $i$-$th$ $(1\leq i \leq n)$ record (tuple) in the $Table$, e.g., Figure \ref{fig06} (b) shows an initial table $R=\{(id1,license1,copy1), (id2,license2,copy2)\}$, we can compute $D_{1}^{2}=license1$. Other operators in DCTL can be obtained through the transformation of the above operators: $deadlock \equiv \neg EXtrue$; $EF\varphi \equiv E(true U \varphi)$; $AF\varphi\equiv  A(true U \varphi)$; $AG\varphi \equiv  \neg EF \neg\varphi$; $AX\varphi \equiv  \neg EX\neg \varphi \wedge \neg deadlock$; and $\varphi_1 \rightarrow \varphi_2 \equiv  \neg\varphi_1 \vee \varphi_2 $.

\begin{definition}[Semantics of DCTL]
	Given an SRG of WFTC-net, $\varphi$ is a DCTL formula. SRG, $c\models \varphi$ means that this formula is satisfied at state c of the SRG. If SRG is clear from the context, it can be omitted. The semantics of relation $\models$ is defined recursively as follows: 
	\begin{itemize}
		\item[(1)] $c\models true$;
		\item[(2)] $c\models p \Leftrightarrow p\in L(c)$;
		\item[(3)] $c\models d \Leftrightarrow d\in L(c)$;
		\item[(4)] $c\models D_{i}^{j} \Leftrightarrow D_{i}^{j} \in L(c)$;
		\item[(5)] $c\models \exists d\in \mathbb{R}, [\varphi \triangle d] \Leftrightarrow \exists d\in \mathbb{R}$ and $c\models \varphi \triangle d$;
		\item[(6)] $c\models \forall d\in \mathbb{R}, [\varphi \triangle d] \Leftrightarrow \forall d\in \mathbb{R}$ and $c\models \varphi \triangle d$;		
		\item[(7)] $c\models \exists D_{i}^{j} \in R, [\varphi \triangle D_{i}^{j}] \Leftrightarrow \exists D_{i}^{j} \in R$ and $c\models \varphi \triangle D_{i}^{j}$;		
		\item[(8)] $c\models \forall D_{i}^{j} \in R, [\varphi \triangle D_{i}^{j}] \Leftrightarrow \forall D_{i}^{j} \in R$ and $c\models \varphi \triangle D_{i}^{j}$;
		\item[(9)] $c \models \neg \varphi  \Leftrightarrow c\not  \models \varphi $;
		\item[(10)] $	c \models {\varphi _1} \vee {\varphi _2} \Leftrightarrow c \models {\varphi _1}$ or $c \models {\varphi _2}$;
		\item[(11)] $	c \models {\rm{AX}}\varphi  \Leftrightarrow \forall (c,{c_1}) \in \Re (c)$ and ${c_1} \models \varphi $; ($\Re(c)$is a binary relation on state $c$)
		\item[(12)] $c \models {\rm{EX}}\varphi  \Leftrightarrow \exists (c,{c_1}) \in \Re (c){\rm{ }}$ and ${\rm{ }}{c_1} \models \varphi $
		\item[(13)] $c \models {\rm{AG}}\varphi  \Leftrightarrow$ for all paths $\langle c_1,c_2 \rangle \langle c_2,c_3 \rangle$..., and all $c_i$ along the path, we have $c_i \models \varphi$, where ${c_1} = c$;
		\item[(14)] $c \models {\rm{EG}}\varphi  \Leftrightarrow$ for some paths  $\langle c_1,c_2 \rangle \langle c_2,c_3 \rangle$..., and all $c_i$ along the path, we have $c_i \models \varphi$, where ${c_1} = c$;
		\item[(15)] $c\models E(\varphi_1 U \varphi_2)\Leftrightarrow$ for some paths $\langle c_1,c_2 \rangle \langle c_2,c_3 \rangle$..., we have: (1) $ \exists k \geq 0, \pi(k) \models \varphi_2$, (2) $\forall 0 \leq j < k, \pi(j) \models \varphi_1$; and
		\item[(16)] $c\models A(\varphi_1 U \varphi_2)\Leftrightarrow$ for all paths $\langle c_1,c_2 \rangle \langle c_2,c_3 \rangle$..., we have: (1) $ \exists k \geq 0, \pi(k) \models \varphi_2$, (2) $\forall 0 \leq j < k, \pi(j) \models \varphi_1$.
	\end{itemize}	
	\label{defn10}
\end{definition}

We now illustrate some properties which can be expressed in DCTL but not succinctly in CTL. A WFTC-net as shown in Figure \ref{fig06} (a), if the system requires that after the user with $id1$ registers driver's license $license1$, then the user with $id2$ cannot register $license1$ again. Traditional CTL cannot describe this specification based on database tables. However, it can be expressed by DCTL formula $\varphi_1$ = $AG(\forall(id1\in R$, $id2\in R)$, $[id1\neq id2 \rightarrow id1.license1\neq id2.license2])$, which means that when the $id1$ and $id2$ are not the same person, their driver's licenses cannot be the same. Model checking algorithm verifies the DCTL formula and calculates the satisfaction set Sat($\bullet$) from the syntax and semantics of DCTL. Algorithm \ref{alg:3} summarizes the basic steps to calculate the satisfaction set.

\begin{algorithm}[tt]
		\caption{Computation of the satisfaction sets}
    	\hspace*{0.02in} {\bf Input:} 	A WFTC-net $N$, SRG, DCTL formula $\varphi$ \\
		\hspace*{0.02in} {\bf Output:} 	Sat($\varphi$)=$\{c\in C| c\models \varphi\}$;
		\begin{algorithmic}[1]
		\If{$\varphi == true$}
		\State Return $(C)$; // $C$ is the set of all states
		\EndIf
		\If{$\varphi == false$}
		\State Return $(\varnothing)$;
		\EndIf
		\If{$\varphi \in p \cup d \cup D_{i}^{j}$}
		\State Return $(\{c| \varphi \in L(c)\})$;
		\EndIf
		\If{$\varphi == \neg \varphi_1$}
		\State Return $(Sat(C-Sat(\varphi_1)))$;
		\EndIf
		\If{$\varphi == \varphi_1 \vee \varphi_2$}
		\State Return $(Sat(\varphi_1) \cup Sat(\varphi_2))$;
		\EndIf
		\If{$\varphi == \varphi_1 \wedge \varphi_2$}
		\State Return $(Sat(\varphi_1) \cap Sat(\varphi_2))$;
		\EndIf
		
		\If{$\varphi == \forall$ $d \in \mathbb{R}, [\varphi \triangle d]$}  
		\For{$c \in C$}
		\If{$\forall$ $c(\varphi) \triangle$ $d$ is true}
		\State Add $c$ into Sat(c);  
		\EndIf
		\EndFor
		\State Return $Sat(c)$;
		\EndIf  
		
		\If{$\varphi == \exists$ $d \in \mathbb{R}, [\varphi \triangle d]$ } 
		\For{$c \in C$}
		\If{$\exists$ $c(\varphi) \triangle$ $d$ is true}
		\State Add $c$ into Sat(c);
		\EndIf
		\EndFor
		\State Return $Sat(c)$;
		\EndIf  
		
		\If{$\varphi == \forall$ $D_{i}^{j} \in R, [\varphi \triangle D_{i}^{j}]$}
		\For{$c \in C$}
		\For{$D_{i}^{j} \in R$}
		\If{$\forall$ $c(\varphi) \triangle D_{i}^{j}$ is true}
		\State Add $c$ into $Sat(c)$;
		\EndIf
		\EndFor
		\EndFor
		\State Return $Sat(c)$;
		\EndIf
		\If{$\varphi == \exists$ $D_{i}^{j} \in R, [\varphi \triangle D_{i}^{j}]$}
		\For{$c \in C$}
		\For{$D_{i}^{j} \in R$}
		\If{$\exists$ $c(\varphi) \triangle D_{i}^{j}$ is true}
		\State Add $c$ into $Sat(c)$;
		\EndIf
		\EndFor
		\EndFor
		\State Return $Sat(c)$;
		\EndIf
	
		\If{$\varphi ==EX\varphi_1 $}
		\State Return $(c\in C|(c,c')\in R\wedge c' \in Sat(\varphi_1))$;
		\EndIf
		
		\If{$\varphi ==EG\varphi_1 $}
		\State Return $(Sat_{EG}(\varphi_1))$;
		\EndIf
		
		\If{$\varphi ==E(\varphi_1 \cup \varphi_2) $}
		\State Return $(Sat_{EU}(\varphi_1,\varphi_2))$;
		\EndIf
		
		\If{$\varphi ==A(\varphi_1 \cup \varphi_2) $}
		\State Return $(Sat_{AU}(\varphi_1,\varphi_2))$;
		\EndIf
	\end{algorithmic}
	\label{alg:3}
	\end{algorithm}

We assume that the number of states and transitions in $N$ are $n$ and $k$, respectively. Given that the computation of the satisfaction sets Sat($\varphi$) is a bottom-up traversal over the parse tree of $\varphi$ and thus linear in $|\varphi|$, the time complexity of Algorithm \ref{alg:3} is in $O((n+k).|\varphi|)$.

\subsection{Model checking algorithms of DCTL}  
This section is concerned with DCTL model checking. We give a transition system(TS) and a DCTL formula $\varphi$. Our detection algorithms can determine whether $TS \models \varphi$ is satisfied. In order to verify the correctness of the system, we need to abstract the system into a WFTC-net model. The specification is abstracted into a formula $\varphi$, $C$ is the set of all states, $Post(c)$ is the successors of state $c$, and $Pre(c)$ is the predecessors of $c$. Algorithm~\ref{alg:2} is used to generate its SRG, and $\varphi$ is verified on the SRG. The formula $\varphi$ satisfies the specification on the SRG. It is equivalent to the system satisfying the specification.

The formula $\varphi$ does not contain data item operations. Its Sat($\bullet $) solution methods can refer to the traditional method\cite{ref26,ref45,ref51}. This paper only gives the algorithm for solving propositions with data, i.e., $\varphi_\triangle$, which means that the proposition formula contains data items related to database table operations. Algorithm ~\ref{alg:3} describes the satisfaction set formulas $Sat_{EX}$, $Sat_{EG}$, $Sat_{EU}$ and $Sat_{AU}$, but does not give detailed steps. This section will provide specific algorithm steps for these formulas. The algorithms $Sat_{EX}$, $Sat_{EG}$, $Sat_{EU}$ and $Sat_{AU}$ of temporal operators $EX$, $EG$, $EU$ and $AU$ are described in Algorithms~\ref{alg:4}, \ref{alg:5}, \ref{alg:6} and \ref{alg:7}, respectively. 

Algorithm~\ref{alg:4} shows the detailed steps for calculating $Sat_{EX}$. Firstly, we calculate the set of states satisfying the formula $\varphi_\triangle$ and store it in a set of $Q_{new}$. Then, we find the predecessors for all states in $Q_{new}$. If the intersection is not empty, we store it in $Q_{old}$. Finally, we find $Sat_{EX}$. 

	\begin{algorithm}[tt]
	\caption{Computation of $Sat_{EX}(\varphi_\triangle)$}
	\hspace*{0.02in} {\bf Input:} 	A WFTC-net $N$, SRG, DCTL formula $Sat_{EX}(\varphi_\triangle)$\\
	\hspace*{0.02in} {\bf Output:} 	$Sat_{EX}(\varphi)$=$\{c\in C| c\models EX \varphi_\triangle \}$
	\begin{algorithmic}[1]
		\State  States $Q_{old}$, $Q_{new}$;
		\For{$ c \in C \wedge c.d \models \varphi_\triangle $} 
		\State  $Q_{new}$=$Sat( \varphi_\triangle)$; //$c.d$ represents the data value of state $c$
		\EndFor
		\State  $Q_{old}$=$\{c\in C| Post(c)\cap Q_{new}\neq \emptyset \}$; // $Post(c)$ is the successors of state $c$
		\State Return $Q_{old}$. 		
	\end{algorithmic}
	\label{alg:4}
\end{algorithm}

For example of the SRG in Figure \ref{fig07}, we can use Algorithm \ref{alg:4} computes the set $Sat_{EX}(\varphi_\triangle)$, e.g., $\varphi1_\triangle$=$EX(id1 \neq id2)$. Our recursive algorithms  obtain the following sets in turns:

1. $Sat(id1)=\{c_0,c_1,...,c_{53}\}$;

2. $Sat(id2)=\{c_0,c_1,...,c_{53}\}$;

3. $Sat(id1 \neq id2)$=$Q_{new}=\{c_0,c_1,...,c_{53}\}$;

4. $Sat(Post(c_i))=\{c_0,c_1,...,c_{52}\}$

5. $Q_{old}$=$Sat(\varphi1_\triangle)$= $Sat(EX(id1 \neq id2))$=$Sat(id1 \neq id2) \cap Sat(Post(c_i))$ =$\{c_0,c_1,...,c_{52}\}$;

Let us consider a computation of $Sat_{EG}(\varphi_\triangle)$, the algorithm for computing $EG(\varphi_\triangle)$ is based on the Definition~\ref{defn10}. Algorithm~\ref{alg:5} shows the detailed steps for calculating $Sat_{EG}(\varphi_\triangle)$. First, we calculate the state set $Q_{old}$ that satisfies the formula $\varphi_\triangle$ and the state set $Q_{new}$ that does not satisfy $\varphi_\triangle$ from the SRG. Then we use the iteration formula $Q_{old}=Sat(\varphi_\triangle)$ and $Q_{old}^{i+1}=Q_{old}^{i}\cap \{c\in Sat(\varphi_\triangle)| Post(c) \cap Q_{old}^{i} \neq \{\varnothing\} \}$ to calculate $Sat_{EG}(\varphi_\triangle)$. Finally, for all $j\geq 0$, it holds that $Q_{old}^{0} {\not \supseteq }$ $Q_{old}^{1}$ ${\not \supseteq }$ $Q_{old}^{2} {\not \supseteq }$ ... ${\not \supseteq }  Q_{old}^{j}=Q_{old}^{j+1}=...=Q_{old}=Sat(\varphi_\triangle)$.

\begin{algorithm}[tt]
	\caption{Computation of $Sat_{EG}(\varphi_\triangle)$}
	\hspace*{0.02in} {\bf Input:} A WFTC-net $N$, SRG, DCTL formula $Sat_{EG}(\varphi_\triangle)$ \\
	\hspace*{0.02in} {\bf Output:} $Sat_{EG}(\varphi_\triangle)$=$\{c\in C| c\models EG \varphi_\triangle \}$
	\begin{algorithmic}[1]
		\State  States $Q_{old}$, $Q_{new}$;
		\For{$ c \in C \wedge c.d \models \varphi_\triangle $} 
		\State  $Q_{old}=Sat(\varphi_\triangle)$; // $Q_{old}$ contains any $c$ for which $c \models EG \varphi_\triangle$ has not been disproven
		\EndFor
		\State  $Q_{new}=C-Sat(\varphi_\triangle)$; // $Q_{new}$ contains any not visited $c'$ with $c' \not  \models EG \varphi_\triangle$ 	
		\For{$\forall c\in Sat(\varphi_\triangle)$}
		\State $Count[c]=|Post(c)|$;
		\EndFor
		\While{$Q_{new}\neq \{\varnothing\}$}
		\State let $c'\in Q_{new}$, $Q_{new}=Q_{new}-{c'}$;  // $Pre(c')$ is the predecessors of state $c'$
		\For{$\forall c\in Pre(c')$} 
		\If{$c \in Q_{old}$}
		\State $Count[c]=Count[c]-1$;
		\If{Count[c]=0}
		\State $Q_{old}=Q_{old}-{c}$;
		\State $Q_{new}=Q_{new}\cup {c}$;
		\EndIf
		\EndIf
		\EndFor
		\EndWhile
		\State Return $Q_{old}$. 
	\end{algorithmic}
	\label{alg:5}
\end{algorithm}

Based on the SRG in Figure \ref{fig07}, we give a formula $\varphi2_\triangle$=$EG(id1 \neq id2)$ and use Algorithms \ref{alg:5} to compute the set $Sat_{EG}(\varphi2_\triangle)$. Our recursive algorithms obtain the following sets in turns:

1. $Sat(id1)=\{c_0,c_1,...,c_{53}\}$;

2. $Sat(id2)=\{c_0,c_1,...,c_{53}\}$;

3. $Sat(id1 \neq id2)$=$Q_{old}=\{c_0,c_1,...,c_{53}\}$;

4. $Q_{new}$ = $C-\{c_0$, $c_1$, ..., $c_{53}\}$ = $\{\varnothing\}$, $count=[3,1,...,0]$;

5. $Q_{old}$=$Sat(\varphi2_\triangle)$=$Sat(EG(id1 \neq id2))$=$\{c_0,c_1,...,c_{53}\}$. 

Algorithm~\ref{alg:6} shows a process of computing $Sat_{EU}(\varphi_{1\triangle}$, $\varphi_{2\triangle})$. It involves computing a minimum fixed point $\cup_i{Q_{new}^{i}}$ of a function $Q_{new}$ = $Sat(\varphi_{2\triangle} \cup$$(Sat(\varphi_{1\triangle})\cap$$\{c\in C| \exists u\in (R(c)\cap Q_{new})\})$. First, we calculate $Q_{new}$, then $Q_{new}(Q_{new})$, and so on, until $Q_{new}^{k}=Q_{new}^{k+1}$, i.e., $Q_{new}^{k+1}=\cup_i Q_{new}^{k}$. Finally, all states contained in $Q_{new}$ satisfy $EU(\varphi_{1\triangle}, \varphi_{2\triangle})$. Due to the finiteness of the SRG and the DCTL formula, the algorithm can be terminated.

\begin{algorithm}[tt]
	\caption{Computation of $Sat_{EU}(\varphi_{1\triangle}, \varphi_{2\triangle})$}
	\hspace*{0.02in} {\bf Input:} 	A WFTC-net $N$, SRG, DCTL formula $Sat_{EU}(\varphi_{1\triangle}, \varphi_{2\triangle})$\\
	\hspace*{0.02in} {\bf Output:} 	$Sat_{EU}(\varphi_{1\triangle}, \varphi_{2\triangle})$=$\{c\in C| c\models EU(\varphi_{1\triangle}, \varphi_{2\triangle})\}$
	\begin{algorithmic}[1]	
		\State  States $Q_{old}$, $Q_{new}$;
		\State  $Q_{old}=\{\varnothing\}$;	
		\For{$ c \in C \wedge c.d \models \varphi_\triangle $} 	
		\State  $Q_{new}=Sat(\varphi_{2\triangle})$;
		\EndFor
		\While{$Q_{old}\neq Q_{new}$}
		\State $Q_{old}=Q_{new}$;
		\State $Q_{new}=Sat(\varphi_{2\triangle})\cup (Sat(\varphi_{1\triangle})\cap \{c|\exists u\in (R(c)\cap Q_{new})\})$; // $u$ represents a non-empty set.
		\EndWhile
		\State Return $Q_{old}$. 
	\end{algorithmic}
	\label{alg:6}
\end{algorithm}

Let us consider the SRG depicted in Figure \ref{fig07}, and suppose we check the formula $Sat_{EU}(\varphi_{1\triangle}, \varphi_{2\triangle})$ with $EU(\varphi_{1\triangle}, \varphi_{2\triangle})$=$EU(id1\neq 0$, $license1\neq 0)$. We invoke Algorithms \ref{alg:6} to compute the set $Sat_{EU}(\varphi_{1\triangle}, \varphi_{2\triangle})$. Our algorithm recursively computes $Sat(EU(id1\neq 0, license1\neq 0))$ as follows: 

1. $Sat(id1)=\{c_0,c_1,...,c_{53}\}$;

2. $Sat(license1)=\{c_0,c_1,...,c_{53}\}$;

3. $Sat(id1 \neq 0)$=$\{c_0,c_1,...,c_{53}\}$;

4. $Sat(license1 \neq 0)$=$\{c_0,c_1,...,c_{53}\}$;

5. $Q_{old}=\{\varnothing\}$, $Q_{new}$=$Sat(license1 \neq 0)$=$\{c_0,c_1,...,c_{53}\}$;

6. $Q_{old}$=$Q_{new}$=$\{c_0,c_1,...,c_{53}\}$; //since $Q_{old}$ $\neq$ $Q_{new}$

7. $Q_{new}$=$Sat(license1 \neq 0)$ $\cup$  $Sat(id1 \neq 0)$ $\cap$ $\{R(c_i) \cup Q_{new}\}$=$\{c_0,c_1,...,c_{53}\}$; // $i\in[0,53]$

Algorithm~\ref{alg:7} shows a process of computing $Sat_{AU}(\varphi_1, \varphi_2)$. It is essentially computing a minimum fixed point $\cup_i Q_{new}^{i}$ of a function $Q_{new}=Sat(\varphi_2) \cup (Sat(\varphi_1)\cap {c\in C| R(c)\subseteq Q_{new}})$. Loop to calculate $Q_{new},Q_{new}(Q_{new})$,..., there exists an integer $k$ such that $Q_{new}^{k}$=$Q_{new}^{k+1}$, i.e., $Q_{new}^{k+1}=\cup_i Q_{new}^{k}$. Finally, all states contained in $Q_{new}$ satisfy $AU(\varphi_1, \varphi_2)$. Due to the finiteness of the SRG and the DCTL formula, the algorithm can be terminated.

\begin{algorithm}[tt]
	\caption{Computation of $Sat_{AU}(\varphi_{1\triangle}, \varphi_{2\triangle})$}
	\hspace*{0.02in} {\bf Input:} A WFTC-net $N$, SRG, DCTL formula $Sat_{AU}(\varphi_1, \varphi_2)$ \\
	\hspace*{0.02in} {\bf Output:} $Sat_{AU}(\varphi_{1\triangle}, \varphi_{2\triangle})$=$\{c\in C| c\models AU(\varphi_{1\triangle}, \varphi_{2\triangle})\}$
	\begin{algorithmic}[1]
		\State  States $Q_{old}$, $Q_{new}$;
		\State  $Q_{old}=\{\varnothing\}$;		
		\For{$ c \in C \wedge c.d \models \varphi_\triangle $} 	
		\State  $Q_{new}=Sat(\varphi_{2\triangle})$;
		\EndFor
		\While{$Q_{old}\neq Q_{new}$}
		\State $Q_{old}=Q_{new}$;
		\State $Q_{new}=Sat(\varphi_{2\triangle})\cup (Sat(\varphi_{1\triangle})\cap \{c| R(c)\subseteq Q_{new})\})$; // $c$ represents a non-empty set
		\EndWhile
		\State Return $Q_{old}$. 
	\end{algorithmic}
	\label{alg:7}
\end{algorithm}

For example of the SRG in Figure \ref{fig07}, we can use Algorithm \ref{alg:7} computes the set $Sat_{AU}(\varphi_{1\triangle}, \varphi_{2\triangle})$, e.g., $AU(\varphi_{1\triangle}$, $\varphi_{2\triangle})$=$AU(id1\neq 0$, $license1\neq 0)$. Our recursive algorithms  obtain the following sets in turns:

1. $Sat(id1)=\{c_0,c_1,...,c_{53}\}$;

2. $Sat(license1)=\{c_0,c_1,...,c_{53}\}$;

3. $Sat(id1 \neq 0)$=$\{c_0,c_1,...,c_{53}\}$;

4. $Sat(license1 \neq 0)$=$\{c_0,c_1,...,c_{53}\}$;

5. $Q_{old}=\{\varnothing\}$, $Q_{new}$=$Sat(license1 \neq 0)$=$\{c_0,c_1,...,c_{53}\}$;

6. $Q_{old}$=$Q_{new}$=$\{c_0,c_1,...,c_{53}\}$; //since $Q_{old}$ $\neq$ $Q_{new}$

7. $Q_{new}$=$Sat(license1 \neq 0)$ $\cup$  $Sat(id1 \neq 0)$ $\cap$ $\{R(c_i) \subseteq Q_{new}\}$=$\{c_0,c_1,...,c_{53}\}$; // $i\in[0,52]$

The time complexity of the CTL model-checking algorithm is determined as follows. Let $n$ be the number of states of $N$ and $k$ be the number of transitions of $N$. Under the assumption that the sets of predecessors Pre(.) are represented as linked lists, the time complexity of Algorithms \ref{alg:4}, Algorithms \ref{alg:5}, Algorithms \ref{alg:6} and Algorithms \ref{alg:7} lies in $O(n+k)$.

\subsection{Model Correctness Verification and Completeness Analysis}   

Model checking can verify an abstract system model. In order to better verify a workflow model under database tables, a DCTL formula verification stage is further refined. The general representation of the DCTL formula is $\varphi=\Gamma$($\Upsilon_1$, $\Upsilon_2$), where the operator $\Gamma$$\in$$\{EX$, $AX$, $EU$, $AU$, $EG$, $AG$, $EF$, $AF\}$. A precondition $\Upsilon_1$ is a quantifier expression involving data items, and a postcondition $\Upsilon_2$ is a combination of one or more propositional functions and logical connectors. Given a WFTC-net $N$, a DCTL formula $\varphi$, and an SRG. Then, the $\Upsilon_1$ and $\Upsilon_2$ are judged in turn. If the $\Upsilon_1$ is $true$ in the SRG, then we calculate $Sat(\Upsilon_1)$ that satisfies $\Upsilon_1$ according to the SRG. Next, we calculate the set $Sat(\Upsilon_2)$ that satisfies $\Upsilon_2$ according to $Sat(\Upsilon_1)$. If $\Upsilon_1$ is $false$ in the SRG, then $\varphi$ is judged to be $false$, indicating that the system does not satisfy the specification. Finally, it is verified whether the initial state $c_0$ satisfies $Sat(\Upsilon_2)$. If it does, it indicates that the system satisfies the specification. Otherwise, the system does not satisfy the specification. Algorithm~\ref{alg:8} gives detailed verification steps.

\begin{algorithm}[htbp][tt]
	\caption{DCTL Verifies the Model Correctness}
	\hspace*{0.02in} {\bf Input:} A WFTC-net $N$, SRG, DCTL formula $\varphi=\Gamma$($\Upsilon_1$,$\Upsilon_2$) \\
	\hspace*{0.02in} {\bf Output:} $N$ $\vDash$ $\varphi$ is $true$, $N$ $\nvDash$ $\varphi$ is $false$
	\begin{algorithmic}[1]
		\State  According to Algorithm~\ref{alg:2}, the SRG of WFTC-net is generated;
		\State   According to Algorithm~\ref{alg:3}, calculate $Sat(\Upsilon_1)$ on SRG;
		\If{$Sat(\Upsilon_1)$$\neq \{\varnothing\}$}
		\State  According to Algorithm~\ref{alg:3}, calculate the $Sat(\Upsilon_2)$ that  satisfies $\Upsilon_2$ from $Sat(\Upsilon_1)$;
		\If{$c_0 \in Sat(\Upsilon_2)$}
		\State  $\varphi \models SRG$;
		\Return $true$;
		\Else
		\State $\varphi \not  \models SRG$;
		\Return $false$;
		\EndIf
		\Else
		\State Return $false$.
		\EndIf
	\end{algorithmic}
	\label{alg:8}
\end{algorithm}

We use DCTL algorithms to verify the following formulas:

$\varphi_1$ = $AG((\forall id1 \in R$, $\forall id2 \in R)$, $[id1\neq id2 \rightarrow id1.license1\neq id2.license2])$; and

$\varphi_2 = EG((\forall id10 \in R), [id10.copy = true])$.

We first verify $\varphi_1$, its equivalent form is $\neg$$ EF(\forall(id1\in R, id2\in R)$, $[(id1\neq id2)\wedge \neg (id1.license1\neq id2.license2)])$. Based on the SRG in Figure~\ref{fig07}, we calculate the pre-formula $\Upsilon_1$ and post-formula $\Upsilon_2$, respectively. Our recursive algorithms obtain the following sets in turns:\\
1. Compute pre-formula to get $Sat(\Upsilon_1)=Sat(\forall id1\in R, \forall id2\in R)=\{c_0,c_1,...,c_{53}\}$;\\
2. Compute post-formula to get $Sat(\Upsilon_2)=Sat([(id1\neq id2) \wedge \neg(id1.license1\neq id2.license2)])$;\\
3. $Sat(id1\neq id2)=\{c_0,c_1,...,c_{53}\}$;\\
4. $Sat(id1.license1\neq id2.license2)=\{c_0,c_1,...,c_{53}\}$;\\
5. $Sat(\neg (id1.license1\neq id2.license2))=\{\oslash \}$;\\
6. $Sat([(id1\neq id2) \wedge \neg (id1.license1 \neq id2.license2)])=\{\oslash \}$;\\
7. $Sat(\varphi_1)=\{c_0,c_1,...,c_{53}\}$;

According to the above analysis, since initial state $c_0 \in Sat(\varphi_1)$ and thus $N\models \varphi_1$.

Similarly, the equivalent form of $\varphi_2$ is $\neg EF(\forall (id10\in R)$,$[\neg (id10.copy == true)])$. From the SRG we know that $id10$ does not belong to $R$, and compute pre-formula $Sat(\Upsilon_1)$=$Sat(\forall (id10 \in R))$=\{$\oslash $\}. From Algorithm~\ref{alg:8}, the returned value is $false$. we have that $N\not  \models \varphi_2$.

The time complexity of our model checking algorithm consists of two parts. First, given the model size, including the number $n$ of states and the transition relationship $r$. Second, given the size of the DCTL formula,  $|\varphi|$ represents the number of atomic propositions and operators. The time complexity of Algorithm~\ref{alg:8} is $O((n+r)\cdot |\varphi|)$.

\section{EVALUATION}

\subsection{IMPLEMENTATION}

We have implemented our algorithm to develop a tool named WFTC-net model checker (short: WFTCMC). WFTCMC written in C++ programming language is developed to automatically verify the correctness of a WFTC-net and show an SRG/C. Then, we conduct a series of experiments to validate our approach. We utilize a PC with Intel Core I5-8500 CPU (3.00GHz) and 8.0G memory to do these experiments to illustrate the efficiency of our tool.

\subsection{Benchmarks}

Table \ref{table03} shows the detailed information of our benchmarks (BM1-BM8). All these benchmarks are converted into the corresponding WFD-Net or WFT-Net. From the table, we can find that the scale in each model is different.

In particular, BM1-BM5 are come from BPMN 2.0 workflow models that provide scenarios \cite{ref46,ref34}. BM6 is the Write-ups example that originated from Reference \cite{ref47}, BM7 is the Travel BPEL process example that originated from Reference \cite{ref48}, and BM8 is our motivating case. We consider the operation of data elements in these benchmarks and the operation of data items in the database tables and make experimental comparisons by adding the guards' constraint set in the corresponding workflow model.

\begin{table}[!htbp]
	\caption{Benchmark Statistics}
	\setlength{\tabcolsep}{0.5mm}
	\tiny
	\begin{center}
		\begin{tabular}{|c|c|c|c|c|c|c|c|} \hline
			\multirow{3}{*}{Benchmarks} & \multirow{3}{*}{Description} & \multicolumn{6}{c|}{Model}  \\ \cline{3-8} 
			& & \multicolumn{5}{c|}{WFD} & WFT\\ \cline{3-8} 
			& & \multicolumn{1}{c|}{$|T|$} & \multicolumn{1}{c|}{$|P|$} & \multicolumn{1}{c|}{$|F|$} & \multicolumn{1}{c|}{$|D|$} & \multicolumn{1}{c|}{$|G|$} & $|R|$ \\ \hline	
			
			\multirow{1}{*}{BM1}&\multicolumn{1}{l|}{\begin{tabular}[c]{@{}l@{}}BM1 is an example of withdrawing money from an ATM.\end{tabular}}&  \multirow{1}{*}{16}  &  \multirow{1}{*}{17}  &  \multirow{1}{*}{37}  &  \multirow{1}{*}{3}   &  \multirow{1}{*}{4}  & 2     \\ \hline
			
			\multirow{1}{*}{BM2}&\multicolumn{1}{l|}{\begin{tabular}[c]{@{}l@{}}BM2 is an online system for reviewing product defect reports\\ and replacing unqualified products.\end{tabular}}&  \multicolumn{1}{c|}{22}  & \multicolumn{1}{c|}{18}  & \multicolumn{1}{c|}{46}  & \multicolumn{1}{c|}{5}   & \multicolumn{1}{c|}{4}   & 3        \\ \hline
			
			\multirow{1}{*}{BM3}&\multicolumn{1}{l|}{\begin{tabular}[c]{@{}l@{}}BM3 is an online shopping platform system. Users can enjoy\\ discounts when they purchase products to a certain amount.\end{tabular}}& \multicolumn{1}{c|}{16} & \multicolumn{1}{c|}{14} & \multicolumn{1}{c|}{32}  & \multicolumn{1}{c|}{10}  & \multicolumn{1}{c|}{4}   & 3     \\ \hline
			
			\multirow{1}{*}{BM4}&\multicolumn{1}{l|}{\begin{tabular}[c]{@{}l@{}}BM4 is a human resources system that selects candidates from\\ among applicants.\end{tabular}}&  \multicolumn{1}{c|}{29}  & \multicolumn{1}{c|}{23}  & \multicolumn{1}{c|}{58}  & \multicolumn{1}{c|}{7}  & \multicolumn{1}{c|}{6}   & 3   \\ \hline
			
			\multirow{1}{*}{BM5}&\multicolumn{1}{l|}{\begin{tabular}[c]{@{}l@{}}BM5 is an example of school student registration management\\ system.\end{tabular}}& \multicolumn{1}{c|}{14}  & \multicolumn{1}{c|}{13}  & \multicolumn{1}{c|}{29}  & \multicolumn{1}{c|}{6}   & \multicolumn{1}{c|}{4}   & 2     \\ \hline
			
			\multirow{1}{*}{BM6}&\multicolumn{1}{l|}{\begin{tabular}[c]{@{}l@{}}BM6 is an example of a company seeks write-ups every month\\ from a selected set of employees for publication in its website.\end{tabular}}& \multicolumn{1}{c|}{20}  & \multicolumn{1}{c|}{17}  & \multicolumn{1}{c|}{40}  & \multicolumn{1}{c|}{5}   & \multicolumn{1}{c|}{6}   & 3    \\ \hline
			
			\multirow{1}{*}{BM7}&\multicolumn{1}{l|}{\begin{tabular}[c]{@{}l@{}}BM7 presents a Travel BPEL process that queries two airlines to\\ identify the ticket with the lower price.\end{tabular}}& \multicolumn{1}{c|}{17}  & \multicolumn{1}{c|}{16}  & \multicolumn{1}{c|}{34}  & \multicolumn{1}{c|}{6}   & \multicolumn{1}{c|}{4}   & 3     \\ \hline
			
			\multirow{1}{*}{BM8}&\multicolumn{1}{l|}{\begin{tabular}[c]{@{}l@{}}BM8 is our motivating example.\end{tabular}}& \multicolumn{1}{c|}{19} & \multicolumn{1}{c|}{14} & \multicolumn{1}{c|}{38}  & \multicolumn{1}{c|}{3}   & \multicolumn{1}{c|}{6}   & 3     \\ \hline
		\end{tabular}
	\end{center}
	\label{table03}
\end{table}

\subsection{Effectiveness of Constraints}
\begin{figure}[tt]
	\centering
	\includegraphics[width=0.44\textwidth]{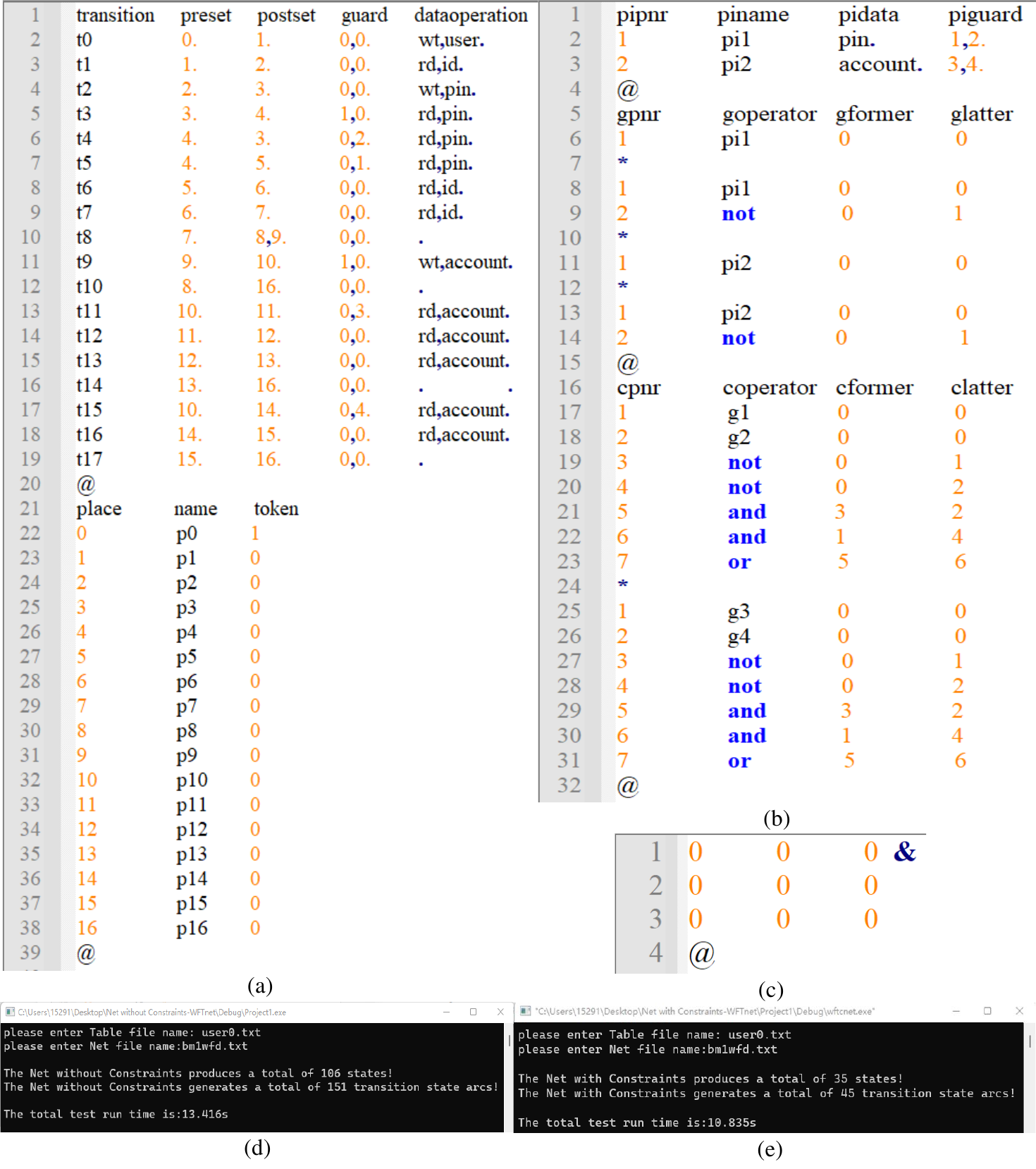}
	\centering
	\caption{(a) A WFD-net; (b) A constraint set; (c) An initial table; (d) SRG; (e) SRGC.}
	\label{fig08}
\end{figure}

We add restrictions to WFT-net and WFD-net for experiments, respectively, and show the effectiveness of constraints through experiments. To describe information in database tables and workflow models. We need to store this information in a .txt file ( When a table information in WFTC-net and data item operations related to table operations are empty, it is a WFD-net). After inputting a WFD-net described in a .txt file, our tool can read them, then generate an SRG, several arcs, and the construction time of the SRG. Figure~\ref{fig08} (a) shows a WFD-net of BM1 in the benchmark experiment; (b) shows predicate functions, guards, and guards constraint sets; (c) shows an initial empty table; (d) shows the results of SRG; (e) shows the results of State Reachability Graph with Constraints (SRGC). From the experimental results in Table \ref{table3}, it is found that when the WFD-net does not add guard constraints, a total of 106 states and 151 arcs are generated. When we add guard constraints to the WFD-net, i.e., WFDC-net, it generates 35 states and 45 arcs. Our method effectively removes 71 pseudo states. It effectively reduces the state space to a certain extent.

\begin{figure}[tt]
	\centering
	\includegraphics[width=0.45\textwidth]{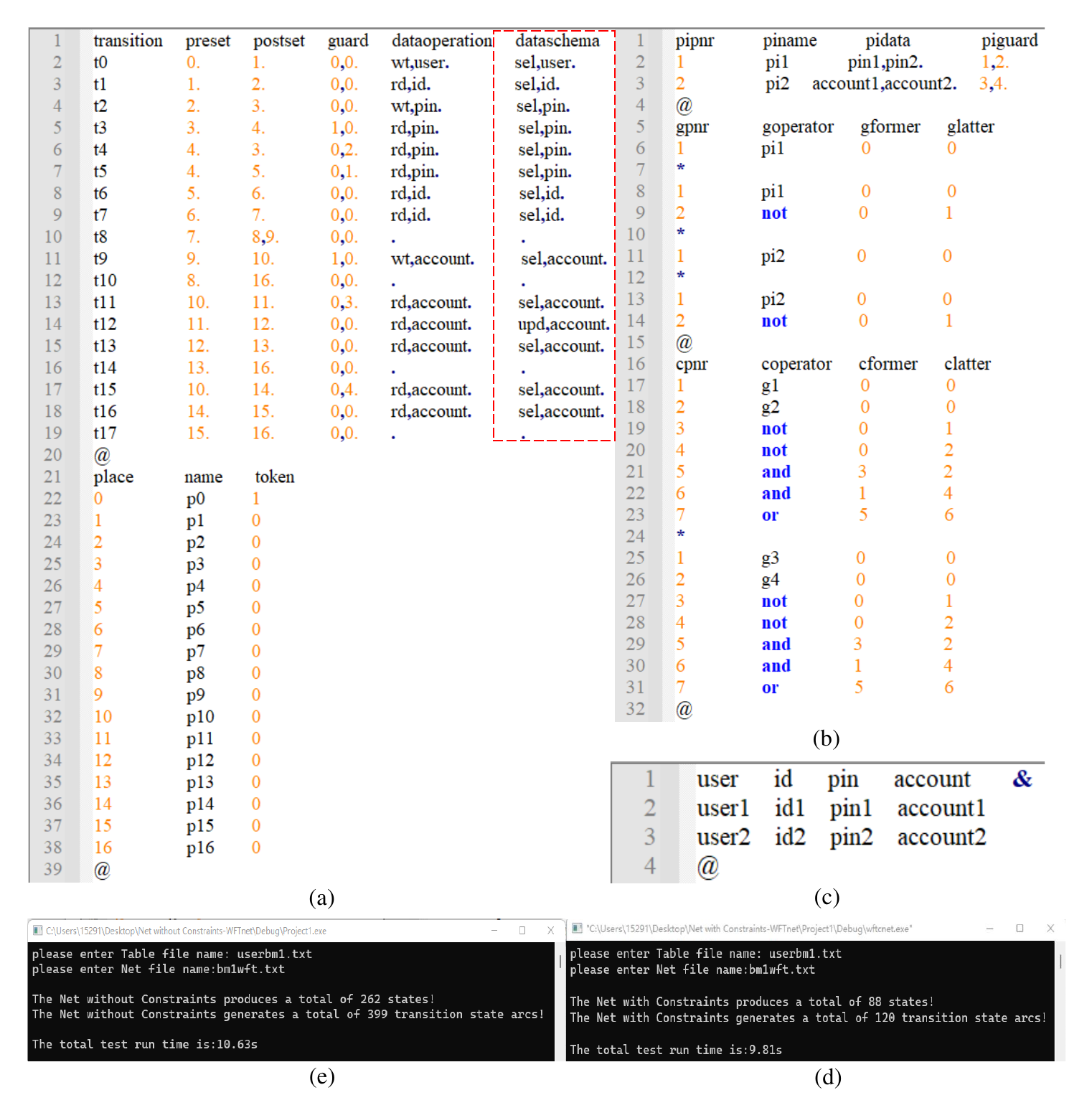}
	\centering
	\caption{(a) A WFT-net; (b) A constraint set; (c) An initial table; (d) SRG; (e) SRGC.}
	\label{fig09}
\end{figure}

We do a similar experimental analysis for a WFT-net, Figure~\ref{fig09} (a) shows a BM1 modeled as the WFT-net (The red dotted box represents operations on data items in the table); (b) shows predicate functions, guards, and guards constraint sets; (c) shows an initial table; (d) shows the results of SRG; (e) shows the results of SRGC. The experimental results are shown in Table \ref{table3} that when the WFT-net does not bind guard constraints, it generates 262 states and 399 arcs. When the guard functions are added to the WFT-net,i.e., WFTC-net, it generates 88 states and 120 arcs. Compared with WFT-net, WFTC net reduces 174 pseudo states.

We do experiments to compare WFD-RG, WFDC-RG, WFT-RG, and WFTC-RG regarding state space and construction time. Experiments are performed on 8 benchmarks. For each benchmark, we first use WFD-net, WFDC-nets, WFT-net, and WFTC-net to model it, then our tool obtains their SRG and SGRC, respectively. Each benchmark was tested 10 times, and the result of running time is their average. Detailed experimental results of the benchmark are shown in Table \ref{table3}.

\begin{table}[htbp]
	\caption{Experimental results}
	\setlength{\tabcolsep}{0.6mm}
	\tiny
	\begin{center}
		\begin{tabular}{|c|cccccc|cccccc|ccccccc}
			\hline
			\multirow{4}{*}{Benchmarks} &  \multicolumn{6}{c|}{State Reachability Graph without Constraints} & \multicolumn{6}{c|}{State Reachability Graph with Constraints}                                                                                                                                                                                                                                                                                                                                                                                       \\ \cline{2-13} 
			&  \multicolumn{3}{c|}{WFD}       & \multicolumn{3}{c|}{WFT}      & \multicolumn{3}{c|}{WFDC}     & \multicolumn{3}{c|}{WFTC}    \\ \cline{2-13} 
			& \multicolumn{1}{c|}{\begin{tabular}[c]{@{}c@{}}Nos.of\\ states\end{tabular}} & \multicolumn{1}{c|}{\begin{tabular}[c]{@{}c@{}}Nos.of\\ arcs\end{tabular}} & \multicolumn{1}{c|}{\begin{tabular}[c]{@{}c@{}}Time\\ (s)\end{tabular}} & \multicolumn{1}{c|}{\begin{tabular}[c]{@{}c@{}}Nos.of\\ states\end{tabular}} & \multicolumn{1}{c|}{\begin{tabular}[c]{@{}c@{}}Nos.of\\ arcs\end{tabular}} & \begin{tabular}[c]{@{}c@{}}Time\\ (s)\end{tabular} & \multicolumn{1}{c|}{\begin{tabular}[c]{@{}c@{}}Nos.of\\ states\end{tabular}} & \multicolumn{1}{c|}{\begin{tabular}[c]{@{}c@{}}Nos.of\\ arcs\end{tabular}} & \multicolumn{1}{c|}{\begin{tabular}[c]{@{}c@{}}Time\\ (s)\end{tabular}} & \multicolumn{1}{c|}{\begin{tabular}[c]{@{}c@{}}Nos.of\\ states\end{tabular}} & \multicolumn{1}{c|}{\begin{tabular}[c]{@{}c@{}}Nos.of\\ arcs\end{tabular}} & \begin{tabular}[c]{@{}c@{}}Time\\ (s)\end{tabular} \\ \hline
			BM1    & \multicolumn{1}{c|}{106}  & \multicolumn{1}{c|}{151}  & \multicolumn{1}{c|}{23.73}  & \multicolumn{1}{c|}{262}   & \multicolumn{1}{c|}{399}  & 20.637   & \multicolumn{1}{c|}{35}  & \multicolumn{1}{c|}{45}  & \multicolumn{1}{c|}{20.32}   & \multicolumn{1}{c|}{88}   & \multicolumn{1}{c|}{120}    & 13.82      \\ \hline
			BM2   & \multicolumn{1}{c|}{208} & \multicolumn{1}{c|}{539} & \multicolumn{1}{c|}{22.649}  & \multicolumn{1}{c|}{494} & \multicolumn{1}{c|}{1317}  & 33.935  & \multicolumn{1}{c|}{67} & \multicolumn{1}{c|}{165}   & \multicolumn{1}{c|}{18.803}  & \multicolumn{1}{c|}{189}  & \multicolumn{1}{c|}{487}    & 25.68     \\ \hline
			BM3 & \multicolumn{1}{c|}{94} & \multicolumn{1}{c|}{102} & \multicolumn{1}{c|}{16.233}  & \multicolumn{1}{c|}{141} & \multicolumn{1}{c|}{173}   & 19.339  & \multicolumn{1}{c|}{35} & \multicolumn{1}{c|}{35} & \multicolumn{1}{c|}{13.46} & \multicolumn{1}{c|}{68}  & \multicolumn{1}{c|}{80} & 18.23 \\ \hline
			BM4  & \multicolumn{1}{c|}{142} & \multicolumn{1}{c|}{152} & \multicolumn{1}{c|}{19.925} & \multicolumn{1}{c|}{266} & \multicolumn{1}{c|}{304} & 25.309 & \multicolumn{1}{c|}{61} & \multicolumn{1}{c|}{64}  & \multicolumn{1}{c|}{14.069} & \multicolumn{1}{c|}{127} & \multicolumn{1}{c|}{140} & 18.99 \\ \hline
			BM5  & \multicolumn{1}{c|}{83}  & \multicolumn{1}{c|}{107} & \multicolumn{1}{c|}{17.58}  & \multicolumn{1}{c|}{115}  & \multicolumn{1}{c|}{143} & 17.881  & \multicolumn{1}{c|}{28}  & \multicolumn{1}{c|}{29}   & \multicolumn{1}{c|}{15.25}   & \multicolumn{1}{c|}{55}   & \multicolumn{1}{c|}{59} & 21.53  \\ \hline
			
			BM6   & \multicolumn{1}{c|}{114}  & \multicolumn{1}{c|}{191}  & \multicolumn{1}{c|}{13.857} & \multicolumn{1}{c|}{325}  & \multicolumn{1}{c|}{556} & 17.585  & \multicolumn{1}{c|}{31} & \multicolumn{1}{c|}{39}  & \multicolumn{1}{c|}{11.416}    & \multicolumn{1}{c|}{105}  & \multicolumn{1}{c|}{137}  & 14.88 \\ \hline
			
			BM7   & \multicolumn{1}{c|}{73} & \multicolumn{1}{c|}{76} & \multicolumn{1}{c|}{10.494} & \multicolumn{1}{c|}{286}  & \multicolumn{1}{c|}{298}  & 18.67 & \multicolumn{1}{c|}{32} & \multicolumn{1}{c|}{31} & \multicolumn{1}{c|}{8.326} & \multicolumn{1}{c|}{125} & \multicolumn{1}{c|}{124} & 14.39  \\ \hline
			
			BM8 & \multicolumn{1}{c|}{227}  & \multicolumn{1}{c|}{334} & \multicolumn{1}{c|}{16.216} & \multicolumn{1}{c|}{629} & \multicolumn{1}{c|}{976} & 19.291 & \multicolumn{1}{c|}{54} & \multicolumn{1}{c|}{65} & \multicolumn{1}{c|}{10.673}  & \multicolumn{1}{c|}{67}  & \multicolumn{1}{c|}{81} & 10.35 \\ \hline
		\end{tabular}
	\end{center}
	\label{table3}
\end{table}

We use constraints in the workflow nets, e.g., WFD-nets and WFT-nets. These models show that our constraint method is universal and effective. From Table \ref{table3}, we can see that the scale of SRGC is much smaller than SRG. It takes less time to produce the SRGC compared with the SRG.

\subsection{Influence of Database Tables Size on State Space}

In this subsection, we study the influence of the size of data items ($|R|$) in the database tables on an SRG. We take the above benchmark BM1, BM2, BM3, and BM8, increasing the $|R|$ number, and compare the size of WFT-RG and WFTC-RG in the state space and the time to construct the corresponding SRG. For each benchmark, we tested it 10 times, and the result of running time is their average. Table~\ref{table4} shows the results of experiments. It shows the scales (i.e., the number of states and arcs) and the construction time of WFT-RG and WFTC-RG for all benchmarks. From this table, we can see that the scale of WFTC-RG is much smaller than WFT-RG. It spends less time producing a WFTC-RG compared to the corresponding WFT-RG. We found that with the increase of the database tables size, SRG and SRGC will increase. Compared with SRG and SRGC in the same model, we found that the size of SRGC decreases significantly with the increase of the database tables size, which further demonstrates the effectiveness of the constraint method, and shows that our method can effectively reduce the rapid growth of the state space.

\begin{table}[!htbp]
	\centering
	\caption{Experimental results}
	\setlength{\tabcolsep}{1.5mm}
	\tiny
	\begin{center}
		\begin{tabular}{|c|c|c|c|c|c|c|c|}
			\cline{1-8}
			\multirow{4}{*}{Benchmarks} & \multicolumn{1}{c|}{WFT-nets} & \multicolumn{3}{c|}{WFT-RG}      & \multicolumn{3}{c|}{WFTC-RG}   \\ \cline{2-8}
			& \multicolumn{1}{c|}{$|R|$} & \multicolumn{1}{c|}{\begin{tabular}[c]{@{}c@{}}Nos.of\\ states\end{tabular}} & \multicolumn{1}{c|}{\begin{tabular}[c]{@{}c@{}}Nos.of\\ arcs\end{tabular}} & \begin{tabular}[c]{@{}c@{}}Time\\ (s)\end{tabular} & \multicolumn{1}{c|}{\begin{tabular}[c]{@{}c@{}}Nos.of\\ states\end{tabular}} & \multicolumn{1}{c|}{\begin{tabular}[c]{@{}c@{}}Nos.of\\ arcs\end{tabular}} & \begin{tabular}[c]{@{}c@{}}Time\\ (s)\end{tabular}  \\ \cline{1-8}
			
			\multirow{3}{*}{BM1}    & \multicolumn{1}{c|}{3}   & \multicolumn{1}{c|}{349}     & \multicolumn{1}{c|}{532}     & 18.095   & \multicolumn{1}{c|}{117}   & \multicolumn{1}{c|}{160}    & 15.662    \\ \cline{2-2} \cline{3-8}
			& \multicolumn{1}{c|}{6}    & \multicolumn{1}{c|}{610}  & \multicolumn{1}{c|}{931}   & 23.153 & \multicolumn{1}{c|}{204}   & \multicolumn{1}{c|}{280}    & 18.699   \\ \cline{2-2} \cline{3-8}
			& \multicolumn{1}{c|}{9}    & \multicolumn{1}{c|}{871} & \multicolumn{1}{c|}{1330}  & 27.943   & \multicolumn{1}{c|}{291}   & \multicolumn{1}{c|}{400}  & 20.697  \\ \cline{1-8}
			
			\multirow{3}{*}{BM2}   & \multicolumn{1}{c|}{3}   & \multicolumn{1}{c|}{494}  & \multicolumn{1}{c|}{1317}  & 20.935  & \multicolumn{1}{c|}{189}   & \multicolumn{1}{c|}{487}  & 17.312   \\ \cline{2-2} \cline{3-8}
			& \multicolumn{1}{c|}{6}    & \multicolumn{1}{c|}{947}    & \multicolumn{1}{c|}{2562}   & 26.368    & \multicolumn{1}{c|}{372}   & \multicolumn{1}{c|}{970}   & 20.293    \\ \cline{2-2} \cline{3-8}
			& \multicolumn{1}{c|}{9}    & \multicolumn{1}{c|}{1400}  & \multicolumn{1}{c|}{3807}    & 35.63   & \multicolumn{1}{c|}{555}  & \multicolumn{1}{c|}{1453}  & 23.223  \\ \cline{1-8}
			
			\multirow{3}{*}{BM3}  & \multicolumn{1}{c|}{3}  & \multicolumn{1}{c|}{141}   & \multicolumn{1}{c|}{173}  & 18.339  & \multicolumn{1}{c|}{68}  & \multicolumn{1}{c|}{80}   & 11.432  \\ \cline{2-2} \cline{3-8}
			& \multicolumn{1}{c|}{6}   & \multicolumn{1}{c|}{201}  & \multicolumn{1}{c|}{260}  & 22.401   & \multicolumn{1}{c|}{107}  & \multicolumn{1}{c|}{125}   & 15.306   \\ \cline{2-2} \cline{3-8}
			& \multicolumn{1}{c|}{9}   & \multicolumn{1}{c|}{261}   & \multicolumn{1}{c|}{347}  & 27.532  & \multicolumn{1}{c|}{134}   & \multicolumn{1}{c|}{170}    & 18.502  \\ \cline{1-8}
			
			\multirow{3}{*}{BM8} & \multicolumn{1}{c|}{3}  & \multicolumn{1}{c|}{629} & \multicolumn{1}{c|}{976} & 19.291  & \multicolumn{1}{c|}{67}   & \multicolumn{1}{c|}{81}   & 10.35   \\ \cline{2-2} \cline{3-8}
			& \multicolumn{1}{c|}{6}   & \multicolumn{1}{c|}{1055}   & \multicolumn{1}{c|}{1672}  & 28.978  & \multicolumn{1}{c|}{159} & \multicolumn{1}{c|}{197}   & 15.454  \\ \cline{2-2} \cline{3-8}
			& \multicolumn{1}{c|}{9}   & \multicolumn{1}{c|}{1481}  & \multicolumn{1}{c|}{2368}    & 34.473  & \multicolumn{1}{c|}{228} & \multicolumn{1}{c|}{284}  & 18.407  \\ \cline{1-8}
		\end{tabular}
	\end{center}
	\label{table4}
\end{table}

\subsection{Evaluation of Tool  Testing Pressure}

\begin{figure}[tt]
	\centering
	\includegraphics[width=0.42\textwidth]{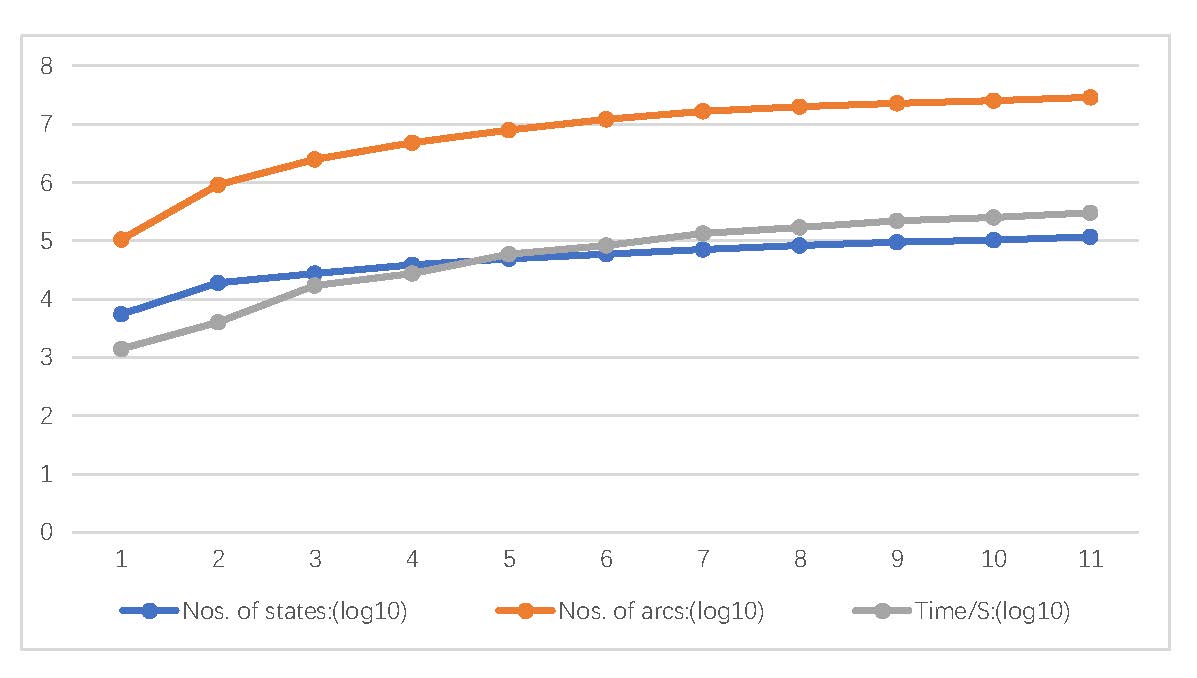}
	\centering
	\caption{The ststes, arcs and running time of SRGC-based methods. All results are obtained by the log10 function.}
	\label{fig10}
\end{figure}

Figure \ref{fig10} is the results of our tool testing pressure on the benchmark BM1, it shows the trend of SRGC's number of states increasing linearly with the table size increase. We expanded the scale of table from 100 to 2100. It shows the scales (i.e., the number of states and arcs) and the construction time of WFTC-RG for BM1. Figure \ref{fig10} shows that the size of an SRGC increases linearly as the scale of the table increases. Especially when the BM1 size reaches 2100, our SRGC can have more than 1.2e+5 states and 3.0e+7 arcs, and their running time is 3.1e+5 (S). The experimental results show that our tool can also run normally when the tool deals with large-scale database tables. 

\subsection{Effectiveness of DCTL Model Checking Method}
\begin{figure} [tt]
	\centering
	\includegraphics[width=0.45\textwidth]{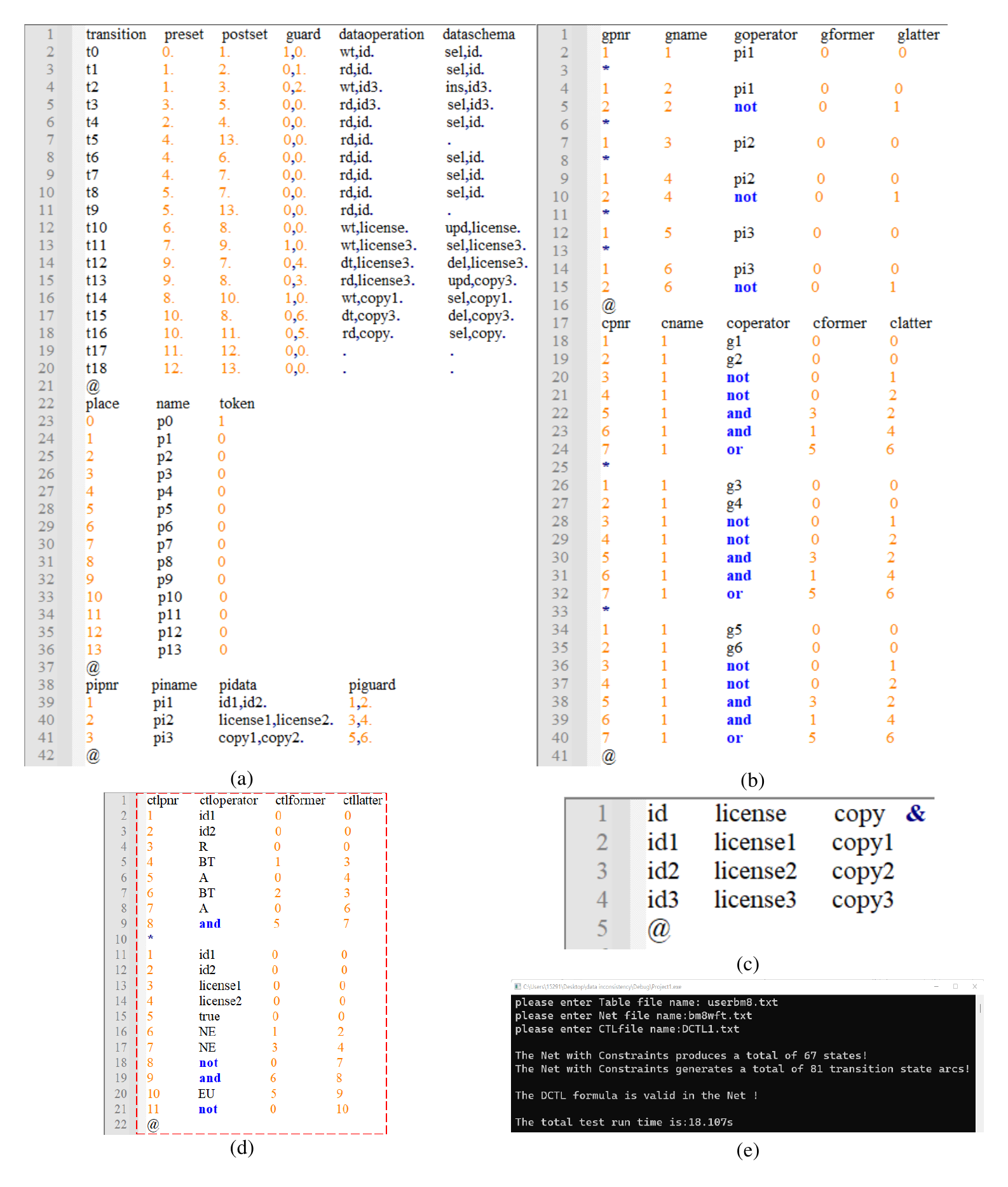}
	\centering
	\caption{(a) A WFTC-net; (b) A constraint set; (c) An initial table; (d) The specification of $\varphi_1$; (e)SRGC.}
	\label{fig11}
\end{figure}

We use the motivation case shown in Figure \ref{fig06} to illustrate the effectiveness of the DCTL model checking method proposed in this paper and use the tool to verify the correctness of the DCTL formula. After inputting a WFTC-net and one DCTL formula  $\varphi_1$ = $AG((\forall id1 \in R$, $\forall id2 \in R)$, $[id1\neq id2 \rightarrow id1.license1\neq id2.license2])$, our tool can output a verification result. Figure~\ref{fig11} (a) Shows a specification of the WFTC-net in Figure ~\ref{fig06}; (b) Shows predicate functions, guards, and guards constraint sets in WFTC-net; (c) Shows an initial table; (d) Shows a specification of DCTL formula; (e) Shows the results of SRGC and verification results; and (f) Shows the results of SRGC. The experimental results show that WFTC-net generates 67 states and 81 arcs. Our tool verifies that the formula $\varphi_1$ is correct.

To further illustrate that DCTL model checking can verify more specifications, Benchmarks (BM1-BM8) are provided in the previous section. Based on the WFTC-net, We give 5 different Behavioural Performance Metrics (PM1-PM5 in Table~\ref{table4}) according to our own needs (researchers can also write their requirements into corresponding Performance Metrics according to their own needs). For each PM, we give the following explanation. PM1: Data items in the same region in the table are increasing (short: increase); PM2: Data items under the different regions in the table are different (short: inequality); PM3: The process model reaches the end (short: terminate); PM4: Check that those data items in the table are not empty (short: non-empty); and PM5: Data items in the detection table are empty (short: empty). At the same time, the specification is given corresponding semantic explanations, and different specifications are described in a DCTL formula, which is illustrated in Figure~\ref{fig06} motivating case. Finally, the tool is used to verify different PM, and the verification results are given. The details of the PM are shown in Table~\ref{table5}.

\begin{table*}[tbp]
	\centering
	\caption{Performance metrics}
	\setlength{\tabcolsep}{1.4mm}
	\tiny
	\begin{center} 
		\begin{tabular}{llll}
			\toprule[1pt]
			\multicolumn{1}{c}{PM}  & \multicolumn{1}{c}{Property}   & \multicolumn{1}{c}{DCTL formula}   &\multicolumn{1}{c}{Example}   \\                                                                                                                                                                          
			\midrule[0.5pt]
			
			\multicolumn{1}{c}{PM1} & \begin{tabular}[c]{@{}l@{}}There exists some reachability paths, \\ the next state of a state satisfies that \\ between   different tuples, the value\\ of the previous data item is smaller\\ than that of the latter.\end{tabular} & \begin{tabular}[c]{@{}l@{}} $\sigma_1$=$EX$(($\forall d1 \in R$, $\forall d2 \in R$), \\$ [d1 \neq d2 \rightarrow d_{1}^{i} < d_{2}^{j}]$)    \end{tabular} & \begin{tabular}[c]{@{}l@{}} $\sigma_1$=$EX$(($\forall id1 \in R, \forall id2 \in R$), \\ $[id1 \neq id2 \rightarrow id1.license1 < id2.license2]$)    	 \end{tabular} \\ \hline
			
			\multicolumn{1}{c}{PM2} & \begin{tabular}[c]{@{}l@{}}In all reachability paths, it is always\\ satisfied that the values of data items \\under the same attribute are different\\ between different tuples.\end{tabular}        & \begin{tabular}[c]{@{}l@{}} $\sigma_2$ = $AG$(($\forall d1 \in R$, $\forall d2 \in R$), \\$[d_1 \neq d_2 \rightarrow d_{1}^{i} \neq d_{2}^{j} ]$)  \end{tabular}      & \begin{tabular}[c]{@{}l@{}} $\sigma_2$=$AG$(($\forall d1 \in R$, $\forall d2 \in R$), \\$[d1\neq d2 \rightarrow  d_{1}^{i} \neq d_{2}^{j}]$)   .\end{tabular}   \\		\hline
			
			\multicolumn{1}{c}{PM3} & \begin{tabular}[c]{@{}l@{}}There exists some reachability paths\\  that eventually reaches the end.\end{tabular}       & \begin{tabular}[c]{@{}l@{}} $\sigma_3$ = $EF$ $P_{end}$    \end{tabular}   & \begin{tabular}[c]{@{}l@{}}      $\sigma_3$ = $EF$ $P_{13}$    \end{tabular}      \\	\hline
			
			\multicolumn{1}{c}{PM4} & \begin{tabular}[c]{@{}l@{}}In all reachability paths, the next state of a \\state satisfies that if the tuple is not empty,\\ then the value of the corresponding data item \\is not empty.\end{tabular}    & \begin{tabular}[c]{@{}l@{}} $\sigma_4$ = $AX$(($\exists d1 \neq \emptyset $), [$d_{1}^{i} \neq \emptyset $]) \end{tabular}          & \begin{tabular}[c]{@{}l@{}}   $\sigma_4$ = $AX$(($\exists id1 \neq \emptyset $), [$id1.copy1 \neq \emptyset $])  \end{tabular}    \\ \hline
			
			\multicolumn{1}{c}{PM5} & \begin{tabular}[c]{@{}l@{}}There exists some reachability path such that\\ the tuple is not empty until the value of a\\ data item corresponding to the tuple is empty.\end{tabular}     & \begin{tabular}[c]{@{}l@{}} $\sigma_5$ = $E$ (($\forall d1 \in R$), [$d1\neq \emptyset$ $U$ $d_{1}^{i} = \emptyset $]) \end{tabular}    & \begin{tabular}[c]{@{}l@{}}  $\sigma_5$ = $E$ (($\forall id3 \in R$), \\ $[id3 \neq \emptyset$ $U$ $id3.license3 = \emptyset ]$)   \end{tabular}      \\							
			\bottomrule[1.3pt]   	   	
		\end{tabular}
	\end{center} 
	\label{table5}
\end{table*}

\begin{table}[htbp]
	\centering
	\caption{ Experimental results of performance metrics}
	\setlength{\tabcolsep}{0.9mm}
	\tiny
	\begin{center}     
			\begin{tabular}{|c|c|c|c|c|c|c|c|}
				\cline{1-8}
				BM      & DCTL & Correct              & Time(s) &             BM        & DCTL & Correct                      & Time(s)  \\ \cline{1-8}
				& PM1  & TRUE                         & 19.326  &                       & PM1  & TRUE                         & 17.955   \\ \cline{2-4} \cline{6-8}
				& PM2  & TRUE                         & 16.855  &                       & PM2  & TRUE                         & 19.603   \\ \cline{2-4} \cline{6-8}
				& PM3  & TRUE                         & 17.084  &                       & PM3  & TRUE                         & 15.919    \\ \cline{2-4} \cline{6-8}
				& PM4  & TRUE                         & 18.795  &                       & PM4  & TRUE                         & 16.622    \\ \cline{2-4} \cline{6-8}
				\multirow{-5}{*}{BM1}& PM5  & TRUE    & 15.946  &           \multirow{-5}{*}{BM5} & PM5  & TRUE             & 15.223    \\ \cline{2-4}  \cline{1-8}
															
				& PM1  & TRUE                         & 18.556  &                       & PM1  & TRUE                         & 16.447    \\ \cline{2-4} \cline{6-8}
				& PM2  & TRUE                         & 16.331  &                       & PM2  & TRUE                         & 18.121   \\ \cline{2-4} \cline{6-8}
				& PM3  & TRUE                         & 19.642  &                       & PM3  & TRUE                         & 13.061    \\ \cline{2-4} \cline{6-8}
				& PM4  & TRUE                         & 18.34   &                       & PM4  & TRUE                         & 15.536  \\ \cline{2-4} \cline{6-8}
				\multirow{-5}{*}{BM2} & PM5 & TRUE    & 14.889   &    \multirow{-5}{*}{BM6} & PM5  & TRUE                     & 18.47   \\ \cline{2-4}  \cline{1-8}				
				
				& PM1  & TRUE                         & 16.32   &                       & PM1  & TRUE                         & 19.149   \\ \cline{2-4} \cline{6-8}
				& PM2  & TRUE                         & 20.226  &                       & PM2  & TRUE                         & 20.336    \\ \cline{2-4} \cline{6-8}
				& PM3  & TRUE                         & 21.896  &                       & PM3  & TRUE                         & 15.516    \\ \cline{2-4} \cline{6-8}
				& PM4  & TRUE                         & 17.005  &                       & PM4  & TRUE                         & 19.239    \\ \cline{2-4} \cline{6-8}
                \multirow{-5}{*}{BM3}& PM5  & TRUE     & 15.977  & \multirow{-5}{*}{BM7} & PM5  & TRUE                         & 14.709    \\ \cline{2-4}   \cline{1-8}
							
				& PM1  & TRUE                         & 29.183  &                       & PM1  & TRUE                         & 13.26     \\ \cline{2-4} \cline{6-8}
				& PM2  & TRUE                         & 21.712  &                       & PM2  & TRUE                         & 15.369   \\ \cline{2-4} \cline{6-8}
				& PM3  & TRUE                         & 26.386  &                       & PM3  & TRUE                         & 14.541    \\ \cline{2-4} \cline{6-8}
				& PM4  & {\textcolor{red} {FALSE}}    & 19.76   &                       & PM4  & TRUE                         & 19.454    \\ \cline{2-4} \cline{6-8}
				\multirow{-5}{*}{BM4}& PM5  & TRUE    & 24.676  & \multirow{-5}{*}{BM8} & PM5  & {\textcolor{red} {FALSE}}    & 18.77     \\ \cline{2-4}  \cline{1-8}							
			\end{tabular}
		\end{center}
		\label{table6} 
	\end{table}

	From the experimental results in Table~\ref{table6}, it can be seen that the results of PM4 in BM4 and PM5 in BM8 are False, and the rest of the results in the table are True. This system error is that in the BM4 model, specification PM4 gives the wrong data $id10$, which does not belong to the data item in the table. BM8 shows the cause of the PM5 error is a user deletes the data item $License$. So the corresponding data item in the table is empty, and the verification result of PM5 is False.

\section{Related Work}

In this section, we review the work related to control flow and data flow in workflow models, as well as existing model checking techniques for analyzing and verifying the correctness of workflow models.

\subsection{Control flow and data flow}
A good modeling method can effectively and quickly analyze the workflow system, and formal method is widely used to verify the correctness of the workflow system. As a formal language, Petri nets are widely used in modeling and analysis of concurrent or distributed systems \cite{ref5, ref24, ref29}, using its structural characteristics to find control flow errors in business process design [44]. Business Process Modeling and Notation (BPMN) and Unified Modeling Language (UML) model the control flow part of business process and analyze its related properties \cite{ref14, ref17}. WFD-net can model data flow and control flow in business processes \cite{ref41}, it combines traditional workflow nets with conceptualized data operations. Each transition binds a guard to represent the impact of the data flow on the activities in the business process \cite{ref34,ref35}. In order to detect possible defects in business process design \cite{ref38}, C. Combi et al. \cite{ref11,ref32} propose an integrated conceptual modeling method of business process and related data, using the concept of activity view to capture the data operations performed by process activities, providing a more rigorous analysis method for business process management. Van Hee et al. \cite{ref54} generate the corresponding model for the execution of database transactions and gave the Petrei nets that generated the execution of these transactions. This model is consistent with transaction execution to some extent. Ghilardi et al. \cite{ref55, ref56} extend the existing coloured Petri nets to generate a network called catalogue and object-aware nets (COA-nets). The transitions are equipped with guards and extracts data from the database through transitions. In order to study the parametric verification problem of COA-nets, this paper points out that it should be coded into a reference framework to solve this problem. Li et al. \cite{ref57} propose a method for modeling event-conditional-action (ECA) rules by conditional colored Petri net (CCPN). This method can rule representation and processing in one model, which is independent of the actual database system. Xiang et al. \cite{ref60} propose a method to transform BPMN into Petri net with Data (PD-net) to detect data flow errors in the model.


The existing studies lack the constraints on guards when formally modeling the workflow system, it's easy to produce Pseudo states. Although some studies have given modeling methods with data operations or with database tables. However, they lack specific description of the state information in each activity and difficult to describe many business relationships in the process. Therefore, some logical defects with data flow errors are difficult to detect in the model. Therefore, we give a modeling method of WFTC-net, it can accurately and carefully describe the workflow system related to database tables operation and more in line with the actual demand.

\subsection{Model Checking}

Model checking is a formal method to automatically verify the correctness of finite state systems. A computer relies on the executed algorithm to search finitely every state in the state space, so as to confirm whether the system model satisfies the system specifications and realize the correctness verification of the business process model \cite{ref2, ref12, ref22,ref52}. In order to use logical expressions to verify the correctness of  process models, E.M. Clarke et al. \cite{ref8, ref9, ref10} give a model checking method, pointing out that model checking is an automated verification technology for finite-state concurrent systems. Jensen J F et al. \cite{ref23} propose a method based on a dependency graph, which introduced a new edge type into the dependency graph and lifted the calculation of fixed points from Boolean domain to a non-negative integer to handle weights. It can be effectively applied to large-scale model checking. H.S. Meda and S. Von Stackelberg et al. \cite{ref26, ref46} propose to use WFD-net to model workflow systems, detect data flow errors in process models by anti-pattern, and use Computation Tree Logic (CTL) to detect process models. Nikola Benes et al. \cite{ref37} consider the problem of parametric synthesis for HCTL properties evaluated over a parametrised Kripke structure (PKS). They give a function that assigns to each of the states of the PKS the set of parametrisations, thus solving the original problem. In order to analyze the modeling and analysis of time-interval specification languages on real-time systems, Chen et al. \cite{ref50} develope a validation system using general theorem prover to facilitate formal analysis of interval-based specification languages and provide machine-aided proof support. Sun and Liu et al. \cite{ref49} propose a language named Stateful Timed CSP and a method for automatically validating stateful timed CSP models capable of specifying a hierarchical real-time system. Through dynamic zone abstraction, finite-state zone graphs can be generated automatically from Stateful Timed CSP models,  and perform model checking.

As workflow systems become increasingly complex, the existing modeling methods and model checking methods are difficult to detect the logic defects in the system, especially the workflow systems related to database table operations. In order to describe and verify the workflow system related to database table operations, we propose a DCTL model checking method, which can describe the specification formula in detail and effectively detect the logic errors in the system.

\section{Conclusion}

In this article, in order to accurately describe a workflow system related to database tables operations, we propose a new model, i.e., WFTC-net; its SRGC can effectively avoid pseudo states and the rapid growth of state space to a certain extent. Our experiments give a good demonstration. On the other hand, in order to verify the logical errors in the workflow model, a DCTL model checking method is proposed. Compared with existing model checking methods, DCTL can describe database table operations and verify logical defects in the workflow model. At the same time, we developed a model checker named WFTCMC. It can automatically generate an SRGC and give a DCTL formula verification result.

In future work, we plan to consider the following research:

(1) We use WFTC nets to detect more data flow errors, such as data loss and data redundancy;

(2) We extend WFTC-nets to consider the deadline of tasks via adding data items with deadlines into transitions, i.e., during the execution of some activities, if the time exceeds the specified time, the system needs to give a corresponding execution strategy. In this way, the system can run normally, and the long wait is avoided;

(3) We consider a time based WFTC-net, transition firing rule, and an SRG generation method;

(4) According to time based WFTC-net, we plan to give its temporal logic expression formulas and verification methods.

\bibliographystyle{IEEEtran}
\bibliography{IEEEabrv,ref.bib}

\begin{thebibliography}{10}
\providecommand{\url}[1]{#1}
\csname url@samestyle\endcsname
\providecommand{\newblock}{\relax}
\providecommand{\bibinfo}[2]{#2}
\providecommand{\BIBentrySTDinterwordspacing}{\spaceskip=0pt\relax}
\providecommand{\BIBentryALTinterwordstretchfactor}{4}
\providecommand{\BIBentryALTinterwordspacing}{\spaceskip=\fontdimen2\font plus
\BIBentryALTinterwordstretchfactor\fontdimen3\font minus
  \fontdimen4\font\relax}
\providecommand{\BIBforeignlanguage}[2]{{%
\expandafter\ifx\csname l@#1\endcsname\relax
\typeout{** WARNING: IEEEtran.bst: No hyphenation pattern has been}%
\typeout{** loaded for the language `#1'. Using the pattern for}%
\typeout{** the default language instead.}%
\else
\language=\csname l@#1\endcsname
\fi
#2}}
\providecommand{\BIBdecl}{\relax}
\BIBdecl

\bibitem{ref67}
J.~Wang and J.~Wang, ``Real-time adaptive allocation of emergency department
  resources and performance simulation based on stochastic timed petri nets,''
  \emph{IEEE Transactions on Computational Social Systems}, 2023.

\bibitem{ref68}
J.~Wang, ``Patient flow modeling and optimal staffing for emergency
  departments: A petri net approach,'' \emph{IEEE Transactions on Computational
  Social Systems}, 2022.

\bibitem{ref9}
X.~Fu, F.~Wang, X.~Liu, K.~Ji, and P.~Zou, ``Dataflow weaknesses analysis of
  scientific workflow based on fault tree,'' in \emph{2012 Sixth International
  Symposium on Theoretical Aspects of Software Engineering}.\hskip 1em plus
  0.5em minus 0.4em\relax IEEE, 2012, pp. 227--230.

\bibitem{ref64}
C.~Z. Zhang, W.~Song, C.~Y. Tang, and F.~F. Chen, ``Detecting data flow errors
  in bpel processes based on smt solvers,'' \emph{Computer Engineering and
  Design}, 2017.

\bibitem{ref70}
L.~Liang, J.~Fu, H.~Zhu, and D.~Liu, ``Solving the team allocation problem in
  crowdsourcing via group multirole assignment,'' \emph{IEEE Transactions on
  Computational Social Systems}, 2022.

\bibitem{ref10}
F.~Jiang, C.-H. Hsu, and S.~Wang, ``Logistic support architecture with petri
  net design in cloud environment for services and profit optimization,''
  \emph{IEEE Transactions on Services Computing}, vol.~10, no.~6, pp. 879--888,
  2016.

\bibitem{ref11}
F.~Moutinho and L.~Gomes, ``Asynchronous-channels within petri net-based gals
  distributed embedded systems modeling,'' \emph{IEEE Transactions on
  Industrial Informatics}, vol.~10, no.~4, pp. 2024--2033, 2014.

\bibitem{ref69}
J.~Liang, Y.~Tang, R.~Hare, B.~Wu, and F.-Y. Wang, ``A learning-embedded
  attributed petri net to optimize student learning in a serious game,''
  \emph{IEEE Transactions on Computational Social Systems}, 2021.

\bibitem{ref7}
L.~M. Castro and T.~Arts, ``Testing data consistency of data-intensive
  applications using quickcheck,'' \emph{Electronic Notes in Theoretical
  Computer Science}, vol. 271, pp. 41--62, 2011.

\bibitem{ref1}
F.~Zhao, D.~Xiang, G.~Liu, and C.~Jiang, ``A new method for measuring the
  behavioral consistency degree of wf-net systems,'' \emph{IEEE transactions on
  computational social systems}, no.~2, p.~9, 2022.

\bibitem{ref2}
M.~Wang, Z.~Ding, G.~Liu, C.~Jiang, and M.~Zhou, ``Measurement and computation
  of profile similarity of workflow nets based on behavioral relation matrix,''
  \emph{IEEE Transactions on Systems, Man, and Cybernetics: Systems}, pp.
  1--18, 2018.

\bibitem{ref3}
M.~Weidlich, A.~Polyvyanyy, N.~Desai, and J.~Mendling, ``Process compliance
  measurement based on behavioural profiles,'' in \emph{International
  Conference on Advanced Information Systems Engineering}, 2010.

\bibitem{ref12}
R.~Dijkman, J.~Hofstetter, and J.~Koehler, ``Business process model and
  notation,'' \emph{Complete Business Process Handbook}, vol.~95, pp. 433--457,
  2011.

\bibitem{ref13}
M.~Fowler, ``Uml distilled: A brief guide to the standard object modeling
  language, 3/e,'' \emph{British Educational Research Journal}, vol.~37, no.~4,
  p. 735–737, 2004.

\bibitem{ref4}
C.~C. Dolean and R.~Petrusel, ``Data-flow modeling: A survey of issues and
  approaches,'' \emph{Informatica Economica Journal}, vol.~16, no.~4, pp.
  117--130, 2012.

\bibitem{ref5}
S.~Sadiq, M.~Orlowska, W.~Sadiq, and C.~Foulger, ``Data flow and validation in
  workflow modelling,'' in \emph{Proceedings of the 15th Australasian database
  conference-Volume 27}.\hskip 1em plus 0.5em minus 0.4em\relax Citeseer, 2004,
  pp. 207--214.

\bibitem{ref8}
D.~Xiang, G.~Liu, C.~Yan, and C.~Jiang, ``Detecting data-flow errors based on
  petri nets with data operations,'' \emph{IEEE/CAA Journal of Automatica
  Sinica}, 2017.

\bibitem{ref14}
N.~Tr{\v{c}}ka, W.~M. Van~der Aalst, and N.~Sidorova, ``Data-flow
  anti-patterns: Discovering data-flow errors in workflows,'' in
  \emph{International Conference on Advanced Information Systems
  Engineering}.\hskip 1em plus 0.5em minus 0.4em\relax Springer, 2009, pp.
  425--439.

\bibitem{ref15}
N.~Sidorova, C.~Stahl, and N.~Tr{\v{c}}ka, ``Workflow soundness revisited:
  Checking correctness in the presence of data while staying conceptual,'' in
  \emph{International Conference on Advanced Information Systems
  Engineering}.\hskip 1em plus 0.5em minus 0.4em\relax Springer, 2010, pp.
  530--544.

\bibitem{ref16}
N.~Sidorova, C.~Stahl, and N.~Trcka, ``Soundness verification for conceptual
  workflow nets with data: Early detection of errors with the most precision
  possible,'' \emph{Information Systems}, vol.~36, no.~7, pp. 1026--1043, 2011.

\bibitem{ref18}
C.~Combi, B.~Oliboni, M.~Weske, and F.~Zerbato, ``Conceptual modeling of
  inter-dependencies between processes and data,'' in \emph{Proceedings of the
  33rd Annual ACM Symposium on Applied Computing}, 2018, pp. 110--119.

\bibitem{ref20}
A.~Cc, A.~Bo, B.~Mw, and C.~Fz, ``Seamless conceptual modeling of processes
  with transactional and analytical data,'' \emph{Data \& Knowledge
  Engineering}, 2021.

\bibitem{ref33}
H.~S. Meda, A.~K. Sen, and A.~Bagchi, ``Detecting data flow errors in
  workflows: A systematic graph traversal approach,'' in \emph{17th Annual
  Workshop on Information Technolgies \& Systems (WITS) Paper}, 2007.

\bibitem{ref34}
S.~Von~Stackelberg, S.~Putze, J.~M{\"u}lle, and K.~B{\"o}hm, ``Detecting
  data-flow errors in bpmn 2.0,'' \emph{Open Journal of Information Systems
  (OJIS)}, vol.~1, no.~2, pp. 1--19, 2014.

\bibitem{ref21}
J.~Ge, H.~Hu, and J.~Lu, ``Invariant analysis for the task refinement of
  workflow nets,'' in \emph{2006 International Conference on Computational
  Inteligence for Modelling Control and Automation and International Conference
  on Intelligent Agents Web Technologies and International Commerce
  (CIMCA'06)}.\hskip 1em plus 0.5em minus 0.4em\relax IEEE, 2006, pp. 209--209.

\bibitem{ref22}
P.~H.~B. Gardiner and C.~Morgan, ``A single complete rule for data
  refinement,'' \emph{Formal Aspects of Computing}, vol.~5, no.~4, pp.
  367--382, 1993.

\bibitem{ref39}
X.~Tao, G.~Liu, B.~Yang, C.~Yan, and C.~Jiang, ``Workflow nets with tables and
  their soundness,'' \emph{IEEE Transactions on Industrial Informatics},
  vol.~16, no.~3, pp. 1503--1515, 2019.

\bibitem{ref35}
N.~Trecka, W.~van~der Aalst, and N.~Sidorova, ``Workflow completion patterns,''
  in \emph{2009 IEEE International Conference on Automation Science and
  Engineering}.\hskip 1em plus 0.5em minus 0.4em\relax IEEE, 2009, pp. 7--12.

\bibitem{ref29}
E.~M. Clarke, ``Model checking,'' in \emph{International Conference on
  Foundations of Software Technology and Theoretical Computer Science}.\hskip
  1em plus 0.5em minus 0.4em\relax Springer, 1997, pp. 54--56.

\bibitem{ref30}
E.~M. Clarke, E.~A. Emerson, and J.~Sifakis, ``Model checking: algorithmic
  verification and debugging,'' \emph{Communications of the ACM}, vol.~52,
  no.~11, pp. 74--84, 2009.

\bibitem{ref27}
A.~S. Dimovski, A.~Legay, and A.~Wasowski, ``Generalized abstraction-refinement
  for game-based ctl lifted model checking,'' \emph{Theoretical Computer
  Science}, vol. 837, pp. 181--206, 2020.

\bibitem{ref37}
N.~Bene{\v{s}}, L.~Brim, S.~Pastva, and D.~{\v{S}}afr{\'a}nek, ``Parallel
  parameter synthesis algorithm for hybrid ctl,'' \emph{Science of Computer
  Programming}, vol. 185, p. 102321, 2020.

\bibitem{ref61}
L.~He, G.~Liu, and M.~Zhou, ``Petri-net-based model checking for
  privacy-critical multiagent systems,'' \emph{IEEE Transactions on
  Computational Social Systems}, 2022.

\bibitem{ref66}
L.~He and G.~Liu, ``Prioritized time-point-interval petri nets modelling
  multi-processor real-time systems and tctl $ \_ $\{$x$\}$ $,'' \emph{IEEE
  Transactions on Industrial Informatics}, 2022.

\bibitem{ref25}
C.~Zhou and Z.~Chen, ``Model checking workflow net based on petri net,''
  \emph{Wuhan University Journal of Natural Sciences}, vol.~11, no.~5, pp.
  1297--1301, 2006.

\bibitem{ref26}
C.~Baier and J.-P. Katoen, \emph{Principles of model checking}.\hskip 1em plus
  0.5em minus 0.4em\relax MIT press, 2008.

\bibitem{ref36}
L.~He and G.~Liu, ``Petri net based symbolic model checking for computation
  tree logic of knowledge,'' \emph{arXiv preprint arXiv:2012.10126}, 2020.

\bibitem{ref52}
A.~D. Lucia, V.~Deufemia, C.~Gravino, and M.~Risi, ``Detecting the behavior of
  design patterns through model checking and dynamic analysis,'' \emph{ACM
  Transactions on Software Engineering and Methodology (TOSEM)}, vol.~26,
  no.~4, pp. 1--41, 2018.

\bibitem{ref40}
G.~Zhou and Z.~Du, ``Petri nets model of implicit data and control in program
  code,'' \emph{Ruanjian Xuebao/Journal of Software}, vol.~22, no.~12, pp.
  2905--2918, 2011.

\bibitem{ref41}
W.~M. Van~der Aalst, ``The application of petri nets to workflow management,''
  \emph{Journal of circuits, systems, and computers}, vol.~8, no.~01, pp.
  21--66, 1998.

\bibitem{ref42}
W.~M. Van Der~Aalst, K.~M. Van~Hee, A.~H. Ter~Hofstede, N.~Sidorova,
  H.~Verbeek, M.~Voorhoeve, and M.~T. Wynn, ``Soundness of workflow nets:
  classification, decidability, and analysis,'' \emph{Formal aspects of
  computing}, vol.~23, no.~3, pp. 333--363, 2011.

\bibitem{ref43}
E.~Bai, N.~Su, Y.~Liang, L.~Qi, and Y.~Du, ``Method for repairing process
  models with selection structures based on token replay,'' \emph{Computing and
  Informatics}, vol.~40, no.~2, pp. 446--468, 2021.

\bibitem{ref23}
D.~Xiang, G.~Liu, C.~Yan, and C.~Jiang, ``A guard-driven analysis approach of
  workflow net with data,'' \emph{IEEE Transactions on Services Computing},
  vol.~14, no.~6, pp. 1650--1661, 2019.

\bibitem{ref44}
------, ``Detecting data inconsistency based on the unfolding technique of
  petri nets,'' \emph{IEEE Transactions on Industrial Informatics}, vol.~13,
  no.~6, pp. 2995--3005, 2017.

\bibitem{ref17}
W.~M. Van Der~Aalst, ``Workflow verification: Finding control-flow errors using
  petri-net-based techniques,'' in \emph{Business Process Management}.\hskip
  1em plus 0.5em minus 0.4em\relax Springer, 2000, pp. 161--183.

\bibitem{ref32}
J.~F. Jensen, K.~G. Larsen, J.~Srba, and L.~K. Oestergaard, ``Efficient
  model-checking of weighted ctl with upper-bound constraints,''
  \emph{International Journal on Software Tools for Technology Transfer},
  vol.~18, no.~4, pp. 409--426, 2016.

\bibitem{ref45}
Z.~Sbai, A.~Missaoui, K.~Barkaoui, and R.~B. Ayed, ``On the verification of
  business processes by model checking techniques,'' in \emph{2010 2nd
  International Conference on Software Technology and Engineering},
  vol.~1.\hskip 1em plus 0.5em minus 0.4em\relax IEEE, 2010, pp. V1--97.

\bibitem{ref51}
S.~Basu, S.~A. Smolka, and O.~R. Ward, ``Model checking the java meta-locking
  algorithm,'' in \emph{Proceedings Seventh IEEE International Conference and
  Workshop on the Engineering of Computer-Based Systems (ECBS 2000)}.\hskip 1em
  plus 0.5em minus 0.4em\relax IEEE, 2000, pp. 342--350.

\bibitem{ref46}
H.~M{\"o}nnich, J.~Raczkowsky, and H.~W{\"o}rn, ``Model checking for robotic
  guided surgery,'' in \emph{International Conference on Electronic
  Healthcare}.\hskip 1em plus 0.5em minus 0.4em\relax Springer, 2009, pp. 1--4.

\bibitem{ref47}
H.~S. Meda, A.~K. Sen, and A.~Bagchi, ``On detecting data flow errors in
  workflows,'' \emph{Journal of Data and Information Quality (JDIQ)}, vol.~2,
  no.~1, pp. 1--31, 2010.

\bibitem{ref48}
W.~Song, C.~Zhang, and H.-A. Jacobsen, ``An empirical study on data flow bugs
  in business processes,'' \emph{IEEE Transactions on Cloud Computing}, vol.~9,
  no.~1, pp. 88--101, 2018.

\bibitem{ref24}
N.~Tr{\v{c}}ka, ``Workflow data footprints,'' in \emph{International Conference
  on Business Information Systems}.\hskip 1em plus 0.5em minus 0.4em\relax
  Springer, 2010, pp. 218--229.

\bibitem{ref38}
P.~Balbiani and C.~Jean-Fran{\c{c}}ois, ``Computational complexity of
  propositional linear temporal logics based on qualitative spatial or temporal
  reasoning,'' in \emph{International Workshop on Frontiers of Combining
  Systems}.\hskip 1em plus 0.5em minus 0.4em\relax Springer, 2002, pp.
  162--176.

\bibitem{ref54}
K.~M. Van~Hee, N.~Sidorova, M.~Voorhoeve, and J.~M. van derWerf, ``Generation
  of database transactions with petri nets,'' \emph{Fundamenta Informaticae},
  vol.~93, no. 1-3, pp. 171--184, 2009.

\bibitem{ref55}
S.~Ghilardi, A.~Gianola, M.~Montali, and A.~Rivkin, ``Petri nets with
  parameterised data,'' in \emph{International Conference on Business Process
  Management}.\hskip 1em plus 0.5em minus 0.4em\relax Springer, 2020, pp.
  55--74.

\bibitem{ref56}
------, ``Petri net-based object-centric processes with read-only data,''
  \emph{Information Systems}, vol. 107, p. 102011, 2022.

\bibitem{ref57}
X.~Li, J.~M. Medina, and S.~V. Chapa, ``Applying petri nets in active database
  systems,'' \emph{IEEE Transactions on Systems, Man, and Cybernetics, Part C
  (Applications and Reviews)}, vol.~37, no.~4, pp. 482--493, 2007.

\bibitem{ref60}
D.~Xiang, S.~Lin, X.~Wang, and G.~Liu, ``Checking missing-data errors in
  cyber-physical systems based on the merged process of petri nets,''
  \emph{IEEE Transactions on Industrial Informatics}, 2022.

\bibitem{ref50}
C.~Chen, J.~Dong, J.~Sun, and A.~Martin, ``A verification system for
  interval-based specification languages,'' \emph{ACM Transactions on Software
  Engineering and Methodology (TOSEM)}, vol.~19, no.~4, pp. 1--36, 2010.

\bibitem{ref49}
J.~Sun, Y.~Liu, J.~Dong, Y.~Liu, L.~Shi, and {\'E}.~Andr{\'e}, ``Modeling and
  verifying hierarchical real-time systems using stateful timed csp,''
  \emph{ACM Transactions on Software Engineering and Methodology (TOSEM)},
  vol.~22, no.~1, pp. 1--29, 2013.

\end{thebibliography}

\section*{Appendix A}

\begin{table*}[htbp]
	\renewcommand{\arraystretch}{0.85}
	\caption{Concrete state information in the state reachability graph in Figure~\ref{fig04}}
	\setlength{\tabcolsep}{1.5mm}
	\footnotesize
	\centering
	\begin{center}
		\begin{tabular}{|c|c|c|c||c|c|c|c|}
			\cline{1-8}
			c   & m   & $\theta_D$ & s & c & m & $\theta_D$  & $\sigma$  \\ \cline{1-8}
			$c_0$  & $p_0$  & \{$\perp$, $\perp$, $\perp$, $\perp$\} & \{$\perp$, $\perp$, $\perp$, $\perp$, $\perp$, $\perp$\} &  $c_{74}$  & $p_{10}$ & \{id, password, license,copy\}  & \{T,T,F,T,F,F\}    \\ \cline{1-8}
			$c_1$  & $p_1$  & \{id, password, $\perp$, $\perp$\}   & \{T,T,$\perp$, $\perp$,$\perp$, $\perp$\}   &  $c_{75}$  & $p_{10}$ & \{id, password, license,copy\}      & \{T,T,F,T,F,T\}       \\ \cline{1-8}
			$c_2$  & $p_1$  & \{id, password, $\perp$, $\perp$\}     & \{T,F,$\perp$, $\perp$, $\perp$, $\perp$\}   & $c_{76}$  & $p_{10}$ & \{id, password, license,copy\}        & \{T,T,T,T,F,F\}   \\ \cline{1-8}
			$c_3$  & $p_1$  & \{id, password, $\perp$, $\perp$\}    & \{F,F,$\perp$, $\perp$, $\perp$, $\perp$\}   & $c_{77}$  & $p_{10}$ & \{id, password, license,copy\} & \{T,T,T,T,F,T\}  \\ \cline{1-8}
			$c_4$  & $p_1$  & \{id, password, $\perp$, $\perp$\}  & \{F,T,$\perp$, $\perp$, $\perp$, $\perp$\}   & $c_{78}$  & $p_{10}$ & \{id, password, license,copy\}    & \{T,T,T,T,T,F\} \\ \cline{1-8}
			$c_5$  & $p_2$  & \{id, password, $\perp$,$\perp$\}  & \{T,T,$\perp$,$\perp$,$\perp$,$\perp$\}     & $c_{79}$  & $p_{10}$ & \{id, password, license,copy\}  & \{T,T,T,T,T,T\}     \\ \cline{1-8}
			$c_6$  & $p_3$  & \{id, password, $\perp$,$\perp$\}    & \{T,T,$\perp$,$\perp$,$\perp$,$\perp$\}   & $c_{80}$  & $p_{11}$ & \{id, password, license,copy\}    & \{T,T,F,T,T,T\}    \\ \cline{1-8}
			$c_7$  & $p_4$  & \{id, password, $\perp$,$\perp$\}     & \{T,T,$\perp$,$\perp$,$\perp$,$\perp$\}  & $c_{81}$  & $p_{12}$ & \{id, password, license,copy\} & \{T,T,F,T,T,T\}   \\ \cline{1-8}
			$c_8$  & $p_{13}$ & \{id, password, $\perp$,$\perp$\}  & \{T,T,$\perp$,$\perp$,$\perp$,$\perp$\}  & $c_{82}$  & $p_{13}$ & \{id, password, license,copy\}   & \{T,T,F,T,T,T\}  \\ \cline{1-8}
			$c_9$  & $p_6$  & \{id, password, $\perp$,$\perp$\}   & \{T,T,$\perp$,$\perp$,$\perp$,$\perp$\}    & $c_{83}$  & $p_{11}$ & \{id, password, license,copy\}   & \{T,T,F,T,F,T\}     \\ \cline{1-8}
			$c_{10}$ & $p_7$  & \{id, password, $\perp$,$\perp$\}      & \{T,T,$\perp$,$\perp$,$\perp$,$\perp$\}     & $c_{84}$  & $p_{12}$ & \{id, password, license,copy\}     & \{T,T,F,T,F,T\}    \\ \cline{1-8}
			$c_{11}$ & $p_{8}$  & \{id, password, license,$\perp$\}    & \{T,T,$\perp$,$\perp$,$\perp$,$\perp$\}       & $c_{85}$  & $p_{13}$ & \{id, password, license,copy\}    & \{T,T,F,T,F,T\}    \\ \cline{1-8}
			$c_{12}$ &  $p_{10}$ & \{id, password, license,copy\}   & \{T,T,$\perp$,$\perp$,T,T\}   & $c_{86}$  & $p_9$  & \{id, password, license,$\perp$\}    & \{T,F,F,F,$\perp$,$\perp$\}   \\ \cline{1-8}
			$c_{13}$ & $p_{10}$ & \{id, password, license,copy\}    & \{T,T,$\perp$,$\perp$,F,F\}   & $c_{87}$  & $p_9$  & \{id, password, license,$\perp$\}    & \{T,F,T,T,$\perp$,$\perp$\}  \\ \cline{1-8}
			$c_{14}$ & $p_{10}$ & \{id, password, license,copy\}     & \{T,T,$\perp$,$\perp$,F,T\}   & $c_{88}$  & $p_9$  & \{id, password, license,$\perp$\}   & \{T,F,T,F,$\perp$,$\perp$\}   \\ \cline{1-8}
			$c_{15}$ & $p_{10}$ & \{id, password, license,copy\}     & \{T,T,$\perp$,$\perp$,T,F\}   & $c_{89}$  & $p_9$  & \{id, password, license,$\perp$\}  & \{T,F,F,T,$\perp$,$\perp$\}   \\ \cline{1-8}
			$c_{16}$ & $p_{11}$ & \{id, password, license,copy\}    & \{T,T,$\perp$,$\perp$,F,T\}    & $c_{90}$  & $p_8$  & \{id, password, license,$\perp$\}   & \{T,F,T,F,$\perp$,$\perp$\}   \\ \cline{1-8}
			$c_{17}$ & $p_{17}$ & \{id, password, license,copy\}    & \{T,T,$\perp$,$\perp$,F,T\}    & $c_{91}$  & $p_8$  & \{id, password, license,$\perp$\}    & \{T,F,T,T,$\perp$,$\perp$\}  \\ \cline{1-8}
			$c_{18}$ & $p_{18}$ & \{id, password, license,copy\}     & \{T,T,$\perp$,$\perp$,F,T\}    & $c_{92}$  & $p_{10}$ & \{id, password, license,copy\}    & \{T,F,T,F,F,F\}     \\ \cline{1-8}
			$c_{19}$ & $p_{11}$ & \{id, password, license,copy\}   & \{T,T,$\perp$,$\perp$,T,T\}   & $c_{93}$  & $p_{10}$ & \{id, password, license,copy\}  & \{T,F,T,F,T,T\}  \\ \cline{1-8}
			$c_{20}$ & $p_{12}$ & \{id, password, license,copy\}    & \{T,T,$\perp$,$\perp$,T,T\}      & $c_{94}$  & $p_{10}$ & \{id, password, license,copy\}   & \{T,F,T,F,T,F\}    \\ \cline{1-8}
			$c_{21}$ & $p_{13}$ & \{id, password, license,copy\}  & \{T,T,$\perp$,$\perp$,T,T\}    & $c_{95}$  & $p_{10}$ & \{id, password, license,copy\}        & \{T,F,T,F,F,T\}    \\ \cline{1-8}
			$c_{22}$ & $p_{13}$ & \{id, password, $\perp$,$\perp$\}  & \{T,T,$\perp$,$\perp$,$\perp$,$\perp$\} & $c_{96}$  & $p_{11}$ & \{id, password, license,copy\}   & \{T,F,T,F,T,T\}    \\ \cline{1-8}
			$c_{23}$ & $p_5$  & \{id, password, $\perp$,$\perp$\}  & \{T,T,$\perp$,$\perp$,$\perp$,$\perp$\}   & $c_{97}$  & $p_{12}$ & \{id, password, license,copy\}   & \{T,F,T,F,T,T\}    \\ \cline{1-8}
			$c_{24}$ & $p_{7}$  & \{id, password, $\perp$,$\perp$\}   & \{T,T,$\perp$,$\perp$,$\perp$,$\perp$\}  & $c_{98}$  & $p_{13}$ & \{id, password, license,copy\}  & \{T,F,T,F,T,T\}      \\ \cline{1-8}
			$c_{25}$ & $p_9$  & \{id, password, license,$\perp$\}   & \{T,T,T,T,$\perp$,$\perp$\}  & $c_{99}$  & $p_{11}$ & \{id, password, license,copy\}      & \{T,F,T,F,F,T\}    \\ \cline{1-8}
			$c_{26}$ & $p_9$  & \{id, password, license,$\perp$\}   & \{T,T,T,F,$\perp$,$\perp$\}      & $c_{100}$ & $p_{12}$ & \{id, password, license,copy\}   & \{T,F,T,F,F,T\}      \\ \cline{1-8}
			$c_{27}$ & $p_9$  & \{id, password, license,$\perp$\}       & \{T,T,F,T,$\perp$,$\perp$\}      & $c_{101}$ & $p_{13}$ & \{id, password, license,copy\}    & \{T,F,T,F,F,T\}      \\ \cline{1-8}
			$c_{28}$ & $p_8$  & \{id, password, license,$\perp$\}  & \{T,T,T,T,$\perp$,$\perp$\} & $c_{102}$ & $p_{10}$ & \{id, password, license,copy\}    & \{T,F,T,T,F,F\}   \\ \cline{1-8}
			$c_{29}$ & $p_{10}$ & \{id, password, license,copy\}     & \{T,T,T,T,T,T\}        & $c_{103}$ & $p_{10}$ & \{id, password, license,copy\}          & \{T,F,T,T,T,F\}    \\ \cline{1-8}
			$c_{30}$ & $p_{10}$ & \{id, password, license,copy\}    & \{T,T,T,T,F,T\}      & $c_{104}$ & $p_{10}$ & \{id, password, license,copy\}   & \{T,F,T,T,F,T\}      \\ \cline{1-8}
			$c_{31}$ & $p_{10}$ & \{id, password, license,copy\}  & \{T,T,T,T,T,F\}    & $c_{105}$ & $p_{10}$ & \{id, password, license,copy\}   & \{T,F,T,T,T,T\}      \\ \cline{1-8}
			$c_{32}$ & $p_{10}$ & \{id, password, license,copy\}  & \{T,T,T,T,F,F\}      & $c_{106}$ & $p_{11}$ & \{id, password, license,copy\}    & \{T,F,T,T,T,T\}  \\ \cline{1-8}
			$c_{33}$ & $p_{11}$ & \{id, password, license,copy\}   & \{T,T,T,T,T,T\}   & $c_{107}$ & $p_{12}$ & \{id, password, license,copy\}     & \{T,F,T,T,T,T\}     \\ \cline{1-8}
			$c_{34}$ & $p_{12}$ & \{id, password, license,copy\}    & \{T,T,T,T,T,T\}  & $c_{108}$ & $p_{13}$ & \{id, password, license,copy\}     & \{T,F,T,T,T,T\}    \\ \cline{1-8}
			$c_{35}$ & $p_{13}$ & \{id, password, license,copy\}     & \{T,T,T,T,T,T\} & $c_{109}$ & $p_{11}$ & \{id, password, license,copy\}      & \{T,F,T,T,F,T\}   \\ \cline{1-8}
			$c_{36}$ & $p_{11}$ & \{id, password, license,copy\}    & \{T,T,T,T,F,T\}   & $c_{110}$ & $p_{12}$ & \{id, password, license,copy\}     & \{T,F,T,T,F,T\}   \\ \cline{1-8}
			$c_{37}$ & $p_{12}$ & \{id, password, license,copy\}    & \{T,T,T,T,F,T\}   & $c_{111}$ & $p_{13}$ & \{id, password, license,copy\}     & \{T,F,T,T,F,T\}    \\ \cline{1-8}
			$c_{38}$ & $p_{13}$ & \{id, password, license,copy\}    & \{T,T,T,T,F,T\}   & $c_{112}$ & $p_3$  & \{id, password, $\perp$,$\perp$\} & \{T,F,$\perp$,$\perp$,$\perp$,$\perp$\}   \\ \cline{1-8}
			$c_{39}$ & $p_8$  & \{id, password, license,$\perp$\}    & \{T,T,F,T,$\perp$,$\perp$\}    & $c_{113}$ & $p_5$  & \{id, password, $\perp$,$\perp$\} & \{T,F,$\perp$,$\perp$,$\perp$,$\perp$\}   \\ \cline{1-8}
			$c_{40}$ & $p_{10}$ & \{id, password, license,copy\}      & \{T,T,F,T,F,F\}   & $c_{114}$ & $p_{13}$ & \{id, password, $\perp$,$\perp$\} & \{T,F,$\perp$,$\perp$,$\perp$,$\perp$\}   \\ \cline{1-8}
			$c_{41}$ & $p_{10}$ & \{id, password, license,copy\}      & \{T,T,F,T,T,T\}   & $c_{115}$ & $p_7$  & \{id, password, $\perp$,$\perp$\} & \{T,F,$\perp$,$\perp$,$\perp$,$\perp$\}   \\ \cline{1-8}
			$c_{42}$ & $p_{10}$ & \{id, password, license,copy\}        & \{T,T,F,T,T,F\}   & $c_{116}$ & $p_9$  & \{id, password, license,$\perp$\}   & \{T,F,F,F,$\perp$,$\perp$\}       \\ \cline{1-8}
			$c_{43}$ & $p_{10}$ & \{id, password, license,copy\}       & \{T,T,F,T,F,T\}     & $c_{117}$ & $p_9$  & \{id, password, license,$\perp$\}   & \{T,F,T,F,$\perp$,$\perp$\}      \\ \cline{1-8}
			$c_{44}$ & $p_{11}$ & \{id, password, license,copy\}       & \{T,T,F,T,T,T\}    & $c_{118}$ & $p_9$  & \{id, password, license,$\perp$\}     & \{T,F,F,T,$\perp$,$\perp$\}    \\ \cline{1-8}
			$c_{45}$ & $p_{12}$ & \{id, password, license,copy\}        & \{T,T,F,T,T,T\}   & $c_{119}$ & $p_9$  & \{id, password, license,$\perp$\}     & \{T,F,T,T,$\perp$,$\perp$\}    \\ \cline{1-8}
			$c_{46}$ & $p_{13}$ & \{id, password, license,copy\}        & \{T,T,F,T,T,T\}     & $c_{120}$ & $p_8$  & \{id, password, license,$\perp$\}   & \{T,F,F,T,$\perp$,$\perp$\}    \\ \cline{1-8}
			$c_{47}$ & $p_{11}$ & \{id, password, license,copy\}         & \{T,T,F,T,F,T\}    & $c_{121}$ & $p_8$  & \{id, password, license,$\perp$\}    & \{T,F,T,T,$\perp$,$\perp$\}     \\ \cline{1-8}
			$c_{48}$ & $p_{12}$ & \{id, password, license,copy\}         & \{T,T,F,T,F,T\}   & $c_{122}$ & $p_{10}$ & \{id, password, license,copy\}    & \{T,F,F,T,F,F\}     \\ \cline{1-8}
			$c_{49}$ & $p_{13}$ & \{id, password, license,copy\}       & \{T,T,F,T,F,T\}     & $c_{123}$ & $p_{10}$ & \{id, password, license,copy\}    & \{T,F,F,T,T,F\}   \\ \cline{1-8}
			$c_{50}$ & $p_2$  & \{id, password, $\perp$,$\perp$\}     & \{T,F,$\perp$,$\perp$,$\perp$,$\perp$\}   & $c_{124}$ & $p_{10}$ & \{id, password, license,copy\}     & \{T,F,F,T,T,T\}         \\ \cline{1-8}
			$c_{51}$ & $p_4$  & \{id, password, $\perp$,$\perp$\}     & \{T,F,$\perp$,$\perp$,$\perp$,$\perp$\}      & $c_{125}$ & $p_{10}$ & \{id, password, license,copy\}        & \{T,F,F,T,F,T\}    \\ \cline{1-8}
			$c_{52}$ & $p_{13}$ & \{id, password, $\perp$,$\perp$\}     & \{T,F,$\perp$,$\perp$,$\perp$,$\perp$\}     & $c_{126}$ & $p_{11}$ & \{id, password, license,copy\}    & \{T,F,F,T,T,T\}    \\ \cline{1-8}
			$c_{53}$ & $p_6$  & \{id, password, $\perp$,$\perp$\}       & \{T,F,$\perp$,$\perp$,$\perp$,$\perp$\}    & $c_{127}$ & $p_{12}$ & \{id, password, license,copy\}      & \{T,F,F,T,T,T\}      \\ \cline{1-8}
			$c_{54}$ & $p_7$  & \{id, password, $\perp$,$\perp$\}     & \{T,F,$\perp$,$\perp$,$\perp$,$\perp$\}   & $c_{128}$ & $p_{13}$ & \{id, password, license,copy\}     & \{T,F,F,T,T,T\}    \\ \cline{1-8}
			$c_{55}$ & $p_8$  & \{id, password, license,$\perp$\}     & \{T,T,$\perp$,$\perp$,$\perp$,$\perp$\}  & $c_{129}$ & $p_{11}$ & \{id, password, license,copy\}       & \{T,F,F,T,F,T\}     \\ \cline{1-8}
			$c_{56}$ & $p_{10}$ & \{id, password, license,copy\}     & \{T,T,$\perp$,$\perp$,F,F\}  & $c_{130}$ & $p_{12}$ & \{id, password, license,copy\}     & \{T,F,F,T,F,T\}    \\ \cline{1-8}
			$c_{57}$ & $p_{10}$ & \{id, password, license,copy\}     & \{T,T,$\perp$,$\perp$,F,T\}   & $c_{131}$ & $p_{13}$ & \{id, password, license,copy\}    & \{T,F,F,T,F,T\}      \\ \cline{1-8}
			$c_{58}$ & $p_{10}$ & \{id, password, license,copy\}     & \{T,T,$\perp$,$\perp$,F,T\}    & $c_{132}$ & $p_{10}$ & \{id, password, license,copy\}      & \{T,F,T,T,F,F\}      \\ \cline{1-8}
			$c_{59}$ & $p_{10}$ & \{id, password, license,copy\}  & \{T,F,$\perp$,$\perp$,T,T\}  & $c_{133}$ & $p_{10}$ & \{id, password, license,copy\}    & \{T,F,T,T,T,F\}   \\ \cline{1-8}
			$c_{60}$ & $p_{11}$ & \{id, password, license,copy\}    & \{T,F,$\perp$,$\perp$,F,T\}    & $c_{134}$ & $p_{10}$ & \{id, password, license,copy\}     & \{T,F,T,T,T,F\}     \\ \cline{1-8}
			$c_{61}$ & $p_{12}$ & \{id, password, license,copy\}     & \{T,F,$\perp$,$\perp$,F,T\}     & $c_{135}$ & $p_{10}$ & \{id, password, license,copy\}   & \{T,F,T,T,T,T\}   \\ \cline{1-8}
			$c_{62}$ & $p_{13}$ & \{id, password, license,copy\}     & \{T,F,$\perp$,$\perp$,F,T\}        & $c_{136}$ & $p_{11}$ & \{id, password, license,copy\}    & \{T,F,T,T,T,F\}   \\ \cline{1-8}
			$c_{63}$ & $p_{11}$ & \{id, password, license,copy\}      & \{T,F,$\perp$,$\perp$,T,T\}       & $c_{137}$ & $p_{12}$ & \{id, password, license,copy\}    & \{T,F,T,T,T,F\}      \\ \cline{1-8}
			$c_{64}$ & $p_{12}$ & \{id, password, license,copy\}      & \{T,F,$\perp$,$\perp$,T,T\}      & $c_{138}$ & $p_{13}$ & \{id, password, license,copy\}       & \{T,F,T,T,T,F\}        \\ \cline{1-8}
			$c_{65}$ & $p_{13}$ & \{id, password, license,copy\}    & \{T,F,$\perp$,$\perp$,T,T\}     & $c_{139}$ & $p_{11}$ & \{id, password, license,copy\}         & \{T,F,T,T,T,T\}      \\ \cline{1-8}
			$c_{66}$ & $p_9$  & \{id, password, license,$\perp$\}    & \{T,T,T,T,$\perp$,$\perp$\}      & $c_{140}$ & $p_{12}$ & \{id, password, license,copy\}      & \{T,F,T,T,T,T\}      \\ \cline{1-8}
			$c_{67}$ & $p_9$  & \{id, password, license,$\perp$\}      & \{T,T,T,F,$\perp$,$\perp$\}       & $c_{141}$ & $p_{13}$ & \{id, password, license,copy\}       & \{T,F,T,T,T,T\}     \\ \cline{1-8}
			$c_{68}$ & $p_9$  & \{id, password, license,$\perp$\}       & \{T,T,F,F,$\perp$,$\perp$\}     & $c_{142}$ & $p_{11}$ & \{id, password, license,copy\}           & \{T,T,T,T,T,T\}    \\ \cline{1-8}
			$c_{69}$ & $p_9$  & \{id, password, license,$\perp$\}      & \{T,T,F,T,$\perp$,$\perp$\}      & $c_{143}$ & $p_{12}$ & \{id, password, license,copy\}       & \{T,T,T,T,T,T\}       \\ \cline{1-8}
			$c_{70}$ & $p_8$  & \{id, password, license,copy\}        & \{T,T,F,T,$\perp$,$\perp$\}      & $c_{144}$ & $p_{13}$ & \{id, password, license,copy\}       & \{T,T,T,T,T,T\}     \\ \cline{1-8}
			$c_{71}$ & $p_8$  & \{id, password, license,copy\}         & \{T,T,T,T,$\perp$,$\perp$\}         & $c_{145}$ & $p_{11}$ & \{id, password, license,copy\}        & \{T,T,T,T,F,T\}    \\ \cline{1-8}
			$c_{72}$ & $p_{10}$ & \{id, password, license,copy\}     & \{T,T,F,T,T,T\}       & $c_{146}$ & $p_{12}$ & \{id, password, license,copy\}    & \{T,T,T,T,F,T\}     \\ \cline{1-8}
			$c_{73}$ & $p_{10}$ & \{id, password, license,copy\}      & \{T,T,F,T,T,F\}      & $c_{147}$ & $p_{13}$ & \{id, password, license,copy\}       & \{T,T,T,T,F,T\}     \\ \cline{1-8}
		\end{tabular}
	\end{center}
	\label{table1}
\end{table*}

\begin{table*}[htbp]
	\renewcommand{\arraystretch}{0.9}
	\caption{Concrete state information in the state reachability graph in Figure~\ref{fig07}}
	\setlength{\tabcolsep}{1.5mm}
	\footnotesize
	\centering
	\begin{center}		
		\begin{tabular}{|c|c|c|c|c|}
				\cline{1-5}
				c  & m & $\theta_D$ & $\vartheta_R$ & $\sigma$   \\ \cline{1-5}
				$c_0$  & $p_0$ & \{$\perp$,$\perp$,$\perp$\} & \{(id1,license1,copy1),(id2,license2,copy2)\} & \{$\perp$,$\perp$,$\perp$,$\perp$,$\perp$,$\perp$\}  \\ \cline{1-5}
				$c_1$  & $p_1$  &\{id1, password, $\perp$,$\perp$\}  & \{(id1,license1,copy1),(id2,license2,copy2)\}  & \{T,T,$\perp$,$\perp$,$\perp$,$\perp$\}  \\ \cline{1-5}
				$c_2$  & $p_1$  &  \{id2, password, $\perp$,$\perp$\}   &  \{(id1,license1,copy1),(id2,license2,copy2)\}   &  \{T,F,$\perp$,$\perp$,$\perp$,$\perp$\}  \\ \cline{1-5}
				$c_3$  &  $p_1$  &  \{id3, password, $\perp$,$\perp$\}  &  \{(id1,license1,copy1),(id2,license2,copy2)\}  &  \{F,T,$\perp$,$\perp$,$\perp$,$\perp$\}   \\ \cline{1-5}
				$c_4$  &  $p_2$  &  \{id1, password, $\perp$,$\perp$\}   &  \{(id1,license1,copy1),(id2,license2,copy2)\}  &  \{T,F,$\perp$,$\perp$,$\perp$,$\perp$\}  \\ \cline{1-5}
				$c_5$  &  $p_4$  &  \{id1, password, $\perp$,$\perp$\}   &  \{(id1,license1,copy1),(id2,license2,copy2)\}   &  \{T,T,$\perp$,$\perp$,$\perp$,$\perp$\}  \\ \cline{1-5}
				$c_6$  &  $p_{13}$ &  \{id1, password, $\perp$,$\perp$\}  &  \{(id1,license1,copy1),(id2,license2,copy2)\}    &  \{T,F,$\perp$,$\perp$,$\perp$,$\perp$\}  \\ \cline{1-5}
				$c_7$  &  $p_6$  &  \{id1, password, $\perp$,$\perp$\}    &  \{(id1,license1,copy1),(id2,license2,copy2)\}   &  \{T,F,$\perp$,$\perp$,$\perp$,$\perp$\}  \\ \cline{1-5}
				$c_8$  &  $p_7$  &  \{id1, password, $\perp$,$\perp$\}    &  \{(id1,license1,copy1),(id2,license2,copy2)\}    &  \{T,F,$\perp$,$\perp$,$\perp$,$\perp$\}   \\ \cline{1-5}
				$c_9$  &  $p_8$  &  \{id1, password, license1,$\perp$\}  &  \{(id1,license1,copy1),(id2,license2,copy2)\}     &  \{T,F,$\perp$,$\perp$,$\perp$,$\perp$\}     \\ \cline{1-5}
				$c_{10}$ &  $p_{10}$ &  \{id1, password, license1,copy3\}  &  \{(id1,license1,copy1),(id2,license2,copy2)\}   &  \{T,F,$\perp$,$\perp$,T,F\} \\ \cline{1-5}
				$c_{11}$ &  $p_{10}$ &  \{id1, password, license1,copy1\}   &  \{(id1,license1,copy1),(id2,license2,copy2)\}  &  \{T,F,$\perp$,$\perp$,F,T\}   \\ \cline{1-5}
				$c_{12}$ &  $p_{11}$ &  \{id1, password, license1,copy1\}   &  \{(id1,license1,copy1),(id2,license2,copy2)\}  &  \{T,F,$\perp$,$\perp$,F,T\}  \\ \cline{1-5}
				$c_{13}$ &  $p_{12}$ &  \{id1, password, license1,copy1\}  &  \{(id1,license1,copy1),(id2,license2,copy2)\}   &  \{T,F,$\perp$,$\perp$,F,T\}   \\ \cline{1-5}
				$c_{14}$ &  $p_{13}$ &  \{id1, password, license1,copy1\}  &  \{(id1,license1,copy1),(id2,license2,copy2)\}   &  \{T,F,$\perp$,$\perp$,F,T\}   \\ \cline{1-5}
				$c_{15}$ &  $p_9$  &  \{id1, password, license3,$\perp$\}   &  \{(id1,license1,copy1),(id2,license2,copy2)\}  &  \{T,F,T,F,$\perp$,$\perp$\}  \\ \cline{1-5}
				$c_{16}$ &  $p_9$  &  \{id1, password, license1,$\perp$\}    &  \{(id1,license1,copy1),(id2,license2,copy2)\}   &  \{T,F,F,T,$\perp$,$\perp$\} \\ \cline{1-5}
				$c_{17}$ &  $p_8$  &  \{id1, password, license1,$\perp$\}  &  \{(id1,license1,copy1),(id2,license2,copy2)\}   &  \{T,F,F,T,$\perp$,$\perp$\}   \\ \cline{1-5}
				$c_{18}$ &  $p_{10}$ &  \{id1, password, license1,copy3\}  &  \{(id1,license1,copy1),(id2,license2,copy2)\}   &  \{T,F,F,T,T,F\}    \\ \cline{1-5}
				$c_{19}$ &  $p_{10}$ &  \{id1, password, license1,copy1\}   &  \{(id1,license1,copy1),(id2,license2,copy2)\}  &  \{T,F,F,T,F,T\}   \\ \cline{1-5}
				$c_{20}$ &  $p_{11}$&  \{id1, password, license1,copy1\}  &  \{(id1,license1,copy1),(id2,license2,copy2)\}  &  \{T,F,F,T,F,T\}     \\ \cline{1-5}
				$c_{21}$ &  $p_{12}$ &  \{id1, password, license1,copy1\}   &  \{(id1,license1,copy1),(id2,license2,copy2)\}   &  \{T,F,F,T,F,T\}   \\ \cline{1-5}
				$c_{22}$ &  $p_{13}$ &  \{id1, password, license1,copy1\}  &  \{(id1,license1,copy1),(id2,license2,copy2)\}  &  \{T,F,F,T,F,T\}  \\ \cline{1-5}
				$c_{23}$ &  $p_2$  &  \{id2, password, $\perp$,$\perp$\}  &  \{(id1,license1,copy1),(id2,license2,copy2)\}  &  \{T,F,$\perp$,$\perp$,$\perp$,$\perp$\} \\ \cline{1-5}
				$c_{24}$ &  $p_4$  &  \{id2, password, $\perp$,$\perp$\}   &  \{(id1,license1,copy1),(id2,license2,copy2)\}   &  \{T,F,$\perp$,$\perp$,$\perp$,$\perp$\}    \\ \cline{1-5}
				$c_{25}$ &  $p_{13}$ &  \{id2, password, $\perp$,$\perp$\}   &  \{(id1,license1,copy1),(id2,license2,copy2)\}   &  \{T,F,$\perp$,$\perp$,$\perp$,$\perp$\} \\ \cline{1-5}
				$c_{26}$ &  $p_6$  &  \{id2, password, $\perp$,$\perp$\}  &  \{(id1,license1,copy1),(id2,license2,copy2)\}  &  \{T,F,$\perp$,$\perp$,$\perp$,$\perp$\}   \\ \cline{1-5}
				$c_{27}$ &  $p_8$  &  \{id2, password, license2,$\perp$\}  &  \{(id1,license1,copy1),(id2,license2,copy2)\}  &  \{T,F,$\perp$,$\perp$,$\perp$,$\perp$\}    \\ \cline{1-5}
				$c_{28}$ &  $p_{10}$ &  \{id2, password, license2,copy3\}  &  \{(id1,license1,copy1),(id2,license2,copy2)\}   &  \{T,F,$\perp$,$\perp$,T,F\}     \\ \cline{1-5}
				$c_{29}$ &  $p_{10}$ &  \{id2, password, license2,copy2\}   &  \{(id1,license1,copy1),(id2,license2,copy2)\}   &  \{T,F,$\perp$,$\perp$,F,T\}    \\ \cline{1-5}
				$c_{30}$ &  $p_{11}$ &  \{id2, password, license2,copy2\}&  \{(id1,license1,copy1),(id2,license2,copy2)\}&  \{T,F,$\perp$,$\perp$,F,T\}    \\ \cline{1-5}
				$c_{31}$ &  $p_{12}$ &  \{id2, password, license2,copy2\}&  \{(id1,license1,copy1),(id2,license2,copy2)\}&  \{T,F,$\perp$,$\perp$,F,T\}     \\ \cline{1-5}
				$c_{32}$ &  $p_{13}$ &  \{id2, password, license2,copy2\}   &  \{(id1,license1,copy1),(id2,license2,copy2)\}   &  \{T,F,$\perp$,$\perp$,F,T\}     \\ \cline{1-5}
				$c_{33}$ &  $p_7$&  \{id2, password, $\perp$,$\perp$\}&  \{(id1,license1,copy1),(id2,license2,copy2)\}   &  \{T,F,$\perp$,$\perp$,$\perp$,$\perp$\}   \\ \cline{1-5}
				$c_{34}$ &  $p_9$&  \{id2, password, license3,$\perp$\}   &  \{(id1,license1,copy1),(id2,license2,copy2)\}   &  \{T,F,T,F,$\perp$,$\perp$\}   \\ \cline{1-5}
				$c_{35}$ &  $p_3$&  \{id3, password, $\perp$,$\perp$\}   &  \{(id1,license1,copy1),(id2,license2,copy2),(id3,$\perp$,$\perp$)\}   &  \{F,T,$\perp$,$\perp$,$\perp$,$\perp$\}    \\ \cline{1-5}
				$c_{36}$ &  $p_5$&  \{id3, password, $\perp$,$\perp$\}    &  \{(id1,license1,copy1),(id2,license2,copy2),(id3,$\perp$,$\perp$)\}   &  \{F,T,$\perp$,$\perp$,$\perp$,$\perp$\}    \\ \cline{1-5}
				$c_{37}$ &  $p_{13}$ &  \{id3, password, $\perp$,$\perp$\}&  \{(id1,license1,copy1),(id2,license2,copy2),(id3,$\perp$,$\perp$)\}   &  \{F,T,$\perp$,$\perp$,$\perp$,$\perp$\}    \\ \cline{1-5}
				$c_{38}$ &  $p_7$&  \{id3, password, $\perp$,$\perp$\}&  \{(id1,license1,copy1),(id2,license2,copy2),(id3,$\perp$,$\perp$)\}   &  \{F,T,$\perp$,$\perp$,$\perp$,$\perp$\}  \\ \cline{1-5}
				$c_{39}$ &  $p_9$&  \{id3, password, license3,$\perp$\}&  \{(id1,license1,copy1),(id2,license2,copy2),(id3,$\perp$,$\perp$)\}   &  \{F,T,T,F,$\perp$,$\perp$\}   \\ \cline{1-5}	
				$c_{40}$ &  $p_9$&  \{id3, password, license1,$\perp$\}&  \{(id1,license1,copy1),(id2,license2,copy2),(id3,$\perp$,$\perp$)\}   &  \{F,T,F,T,$\perp$,$\perp$\}    \\ \cline{1-5}
				$c_{41}$ &  $p_8$&  \{id3, password, license1,$\perp$\}&  \{(id1,license1,copy1),(id2,license2,copy2),(id3,license1,$\perp$)\}   &  \{F,T,F,T,$\perp$,$\perp$\}   \\ \cline{1-5}
				$c_{42}$ &  $p_{10}$ &  \{id3, password, license1,copy3\}    &  \{(id1,license1,copy1),(id2,license2,copy2),(id3,license1,$\perp$)\}   &  \{F,T,F,T,T,F\}    \\ \cline{1-5}
				$c_{43}$ &  $p_{10}$ &  \{id3, password, license1,copy2\}    &  \{(id1,license1,copy1),(id2,license2,copy2),(id3,$\perp$,$\perp$)\}   &  \{F,T,F,T,F,T\}   \\ \cline{1-5}
				$c_{44}$ &  $p_{11}$ &  \{id3, password, license1,copy2\}   &  \{(id1,license1,copy1),(id2,license2,copy2),(id3,$\perp$,$\perp$)\}   &  \{F,T,F,T,F,T\}    \\ \cline{1-5}
				$c_{45}$ &  $p_{12}$ &  \{id3, password, license1,copy2\}   &  \{(id1,license1,copy1),(id2,license2,copy2),(id3,$\perp$,$\perp$)\}   &  \{F,T,F,T,F,T\}   \\ \cline{1-5}
				$c_{46}$ &  $p_{13}$ &  \{id3, password, license1,copy2\}    &  \{(id1,license1,copy1),(id2,license2,copy2),(id3,$\perp$,$\perp$)\}   &  \{F,T,F,T,F,T\}     \\ \cline{1-5}
				$c_{47}$ &  $p_9$&  \{id2, password, license1,$\perp$\}     &  \{(id1,license1,copy1),(id2,license2,copy2)\}&  \{T,F,F,T,$\perp$,$\perp$\}    \\ \cline{1-5}
				$c_{48}$ &  $p_8$&  \{id2, password, license1,$\perp$\}   &  \{(id1,license1,copy1),(id2,license1,copy2)\}  &  \{T,F,F,T,$\perp$,$\perp$\}  \\ \cline{1-5}
				$c_{49}$ &  $p_{10}$ &  \{id2, password, license1,copy3\}   &  \{(id1,license1,copy1),(id2,license1,copy2)\}   &  \{T,F,F,T,T,F\}   \\ \cline{1-5}
				$c_{50}$ &  $p_{10}$ &  \{id2, password, license1,copy2\}   &  \{(id1,license1,copy1),(id2,license1,copy2)\}   &  \{T,F,F,T,F,T\}   \\ \cline{1-5}
				$c_{51}$ &  $p_{11}$ &  \{id2, password, license1,copy2\}   &  \{(id1,license1,copy1),(id2,license1,copy2)\}   &  \{T,F,F,T,F,T\}   \\ \cline{1-5}
				$c_{52}$ &  $p_{12}$ &  \{id2, password, license1,copy2\}   &  \{(id1,license1,copy1),(id2,license1,copy2)\}  &  \{T,F,F,T,F,T\}   \\ \cline{1-5}
				$c_{53}$ &  $p_{13}$ &  \{id2, password, license1,copy2\}   &  \{(id1,license1,copy1),(id2,license1,copy2)\}   &  \{T,F,F,T,F,T\}  \\ \cline{1-5}
			\end{tabular}
		\end{center}
		\label{table2}
	\end{table*}

	\ifCLASSOPTIONcaptionsoff
	\newpage
	\fi

\end{document}